%% file: main.tex
\documentclass[pdflatex]{sn-jnl}% Math and Physical Sciences Numbered 
\usepackage{graphicx}%
\usepackage{lineno}
\usepackage{multirow}%
\usepackage{amsmath,amssymb,amsfonts}%
\usepackage{amsthm}%
\usepackage{mathrsfs}%
\usepackage[title]{appendix}%
\usepackage{xcolor}%
\usepackage{textcomp}%
\usepackage{manyfoot}%
\usepackage{booktabs}%
\usepackage{algorithm}%
\usepackage{algorithmicx}%
\usepackage{algpseudocode}%
\usepackage{listings}%
\usepackage{array}
\usepackage{natbib}
\usepackage{bm}
\usepackage[caption=false,font=normalsize,labelfont=sf,textfont=sf]{subfig}
\usepackage{textcomp}
\usepackage{stfloats}
\usepackage{url}
\usepackage{verbatim}
\usepackage{graphicx}
\usepackage{diagbox}
\usepackage{lineno}
\newcommand{\tabincell}[2]{\begin{tabular}{@{}#1@{}}#2\end{tabular}}

\theoremstyle{thmstyleone}

\theoremstyle{thmstyletwo}%

\theoremstyle{thmstylethree}%

\begin{document}
\title[Article Title]{Huayu: Advanced Real-Time Precipitation Estimation from Geostationary Satellite} %% Article title

\author[1,2]{\fnm{Zijiang} \sur{Song}}\email{zjsong@stu.ecnu.edu.cn}
\author[1,2]{\fnm{Ting} \sur{Liu}}
\author*[1,2]{\fnm{Lina} \sur{Yuan}}\email{lnyuan@geoai.ecnu.edu.cn}
\author[1,2]{\fnm{Yuying} \sur{Li}}
\author[1,2]{\fnm{Ao} \sur{Xu}}
\author[3]{\fnm{Xigang} \sur{Sun}}
\author*[1,2]{\fnm{Ye} \sur{Li}}\email{yli@geo.ecnu.edu.cn}
\author[4]{\fnm{Feng} \sur{Lu}}
\author*[1,2]{\fnm{Min} \sur{Liu}}\email{mliu@geo.ecnu.edu.cn}

\affil[1]{\orgdiv{Key Laboratory of Geographic Information Science (Ministry of Education)}, \orgname{East China Normal University}, \orgaddress{\city{Shanghai}, \postcode{200241}, \country{China}}}

\affil[2]{\orgdiv{Key Laboratory of Spatial-temporal Big Data Analysis and Application of Natural Resources in Megacities (Ministry of Natural Resources)}, \orgname{East China Normal University}, \orgaddress{\city{Shanghai}, \postcode{200241}, \country{China}}}

\affil[3]{\orgdiv{School of Computer Science and Engineering}, \orgname{Southeast University}, \orgaddress{\city{Nanjing}, \postcode{211189}, \country{China}}}

\affil[4]{\orgname{Key Laboratory of Radiometric Calibration and Validation for Environmental Satellites, National Satellite Meteorological Center (National Center for Space Weather), Innovation Center for Feng Yun Meteorological Satellite (FYSIC), China Meteorological Administration}, \orgaddress{\city{Beijing}, \postcode{100081}, \country{China}}}

\input{content/sections/0.abstract}
\maketitle
\input{content/content}

\end{document}

%% file: content/sections/0.abstract.tex
    \abstract{As climate change drives increased frequency and intensity of extreme precipitation and flooding worldwide, posing escalating threats to public safety and economic assets, accurate and real-time satellite-based precipitation estimation is essential for operational large-scale hydrometeorological analysis and disaster monitoring. NASA’s Integrated Multi-satellitE Retrievals for GPM (IMERG Final Run) combines information from “all” satellite microwave observations with gauge correction and climatological adjustment to produce precipitation estimates at  0.1° spatial and 30-min temporal resolution. However, its latency of approximately 3.5 months restricts its utility for real-time applications, despite outperforming mainstream satellite precipitation datasets in representing rainfall patterns and variability. We present Huayu, a novel machine learning-based real-time satellite precipitation retrieval system that relies solely on infrared observations from the FengYun-4B geostationary satellite to provide a more accurate precipitation estimate at a finer spatiotemporal resolution (15 min, 0.05°) over a 120° × 120° domain. Performance evaluations demonstrate that Huayu achieves strong consistency with rain gauge observations, yielding a critical success index (CSI) of 0.693 - representing a 3.43\% improvement over IMERG Final Run (CSI: 0.670). Experimental results confirm that infrared satellite observations can deliver more accurate precipitation estimates than conventional multi-source algorithms. }

%% file: content/content.tex
\input{content/sections/1.introduction}
\input{content/sections/4.experiments}

\input{content/sections/3.methods}
\input{content/sections/discussion}
\input{content/sections/6.information}

\bibliographystyle{sn-mathphys-num}
\bibliography{content/reference}

\input{content/sections/Supplementary}

%% file: content/sections/1.introduction.tex
\begin{figure}[!htbp]
\centering
\centerline{\includegraphics[width=1\textwidth]{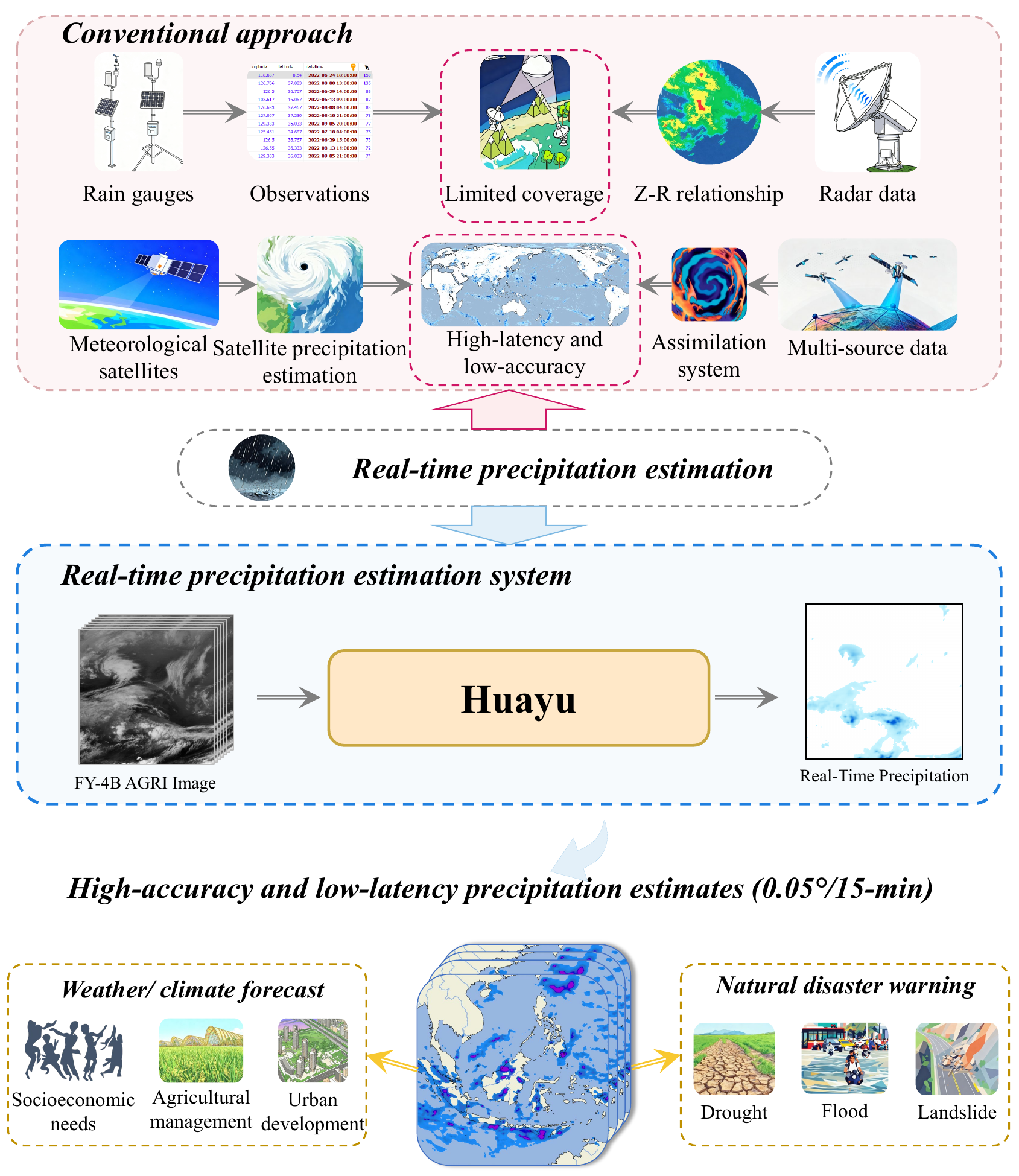}}
\caption{Huayu system versus conventional approaches. The Huayu real-time system provides high-resolution, low-latency precipitation estimates, overcoming the limited coverage and high latency of conventional observation and assimilation methods.}
\label{fig:introduction}
\end{figure}
\section{Introduction}
Precipitation is a critical weather parameter with profound scientific and societal importance. Accurate and real-time data are essential for disaster response, agricultural planning, and water resource management, enabling early warnings for floods and droughts. However, the high variability and intensity of rainfall make its accurate, large-scale estimation particularly challenging, as illustrated in Fig.~\ref{fig:introduction}.

Current real-time precipitation monitoring primarily relies on ground-based weather radar systems equipped with active microwave technology~\citep{prudden2020review}. The most widely employed method for deriving precipitation estimates from weather radar data is the Z-R relationship~\citep{fulton1998wsr}. Recent advancements include machine learning models like NowcastNet, which specializes in predicting extreme rainfall events up to three hours in advance ~\citep{andrychowicz2023deep}. Similarly, Google's DeepMind has utilized the MetNet model within the Multi-Radar/Multi-Sensor System (MRMS) to achieve accurate precipitation predictions up to twelve hours ahead~\citep{espeholt2022deep,sonderby2020metnet,zhang2023skilful}. Despite these improvements, the spatial coverage of radar networks remains a major constraint. This limitation is especially acute in remote and oceanic regions, where scarce radar data impedes comprehensive large-scale precipitation monitoring ~\citep{prudden2020review,wang2020infrared}.

Advances in satellite observation technology have introduced innovative methods for regional precipitation monitoring~\citep{cheng2016survey,gomez2015multimodal,navalgund2007remote,goetz1983remote}. Polar-orbiting satellites fly at lower altitudes and offer higher spatial resolution, but they revisit the same region less frequently. In contrast, geostationary satellites, orbiting at approximately 36,000 km, deliver consistent, high-frequency observations of a fixed area.  

The rapid evolution of machine learning has significantly influenced meteorological satellite research. Using satellite-based machine learning to produce precipitation estimation has been proven to be a feasible solution. Compared with the classic methods~\citep{ji2025new,kubota2020global,tan2019imerg,zhu2024pecam}, machine learning could achieve a more advanced performance and build a more complex relationship between satellite precipitation data with less auxiliary data. \cite{wang2020infrared} developed IPEC (Infrared Precipitation Estimation using a Convolutional neural network) using GOES IR channels (bands 3/4/6). When validated against the NCEP Stage-IV precipitation analysis  ~\citep{lin2011gcip}, the model achieved a Critical Success Index (CSI) of approximately 0.4. In a subsequent study,   \cite{wang2021infrared} adapted the framework for FengYun satellites to cover China, creating IPEC-V2. When evaluated on the CMPA gridded precipitation product ~\citep{shen2014high}, IPEC-V2 attained a maximum Correlation Coefficient (CC) of 0.34.  \cite{jiang2023mcspf} developed a precipitation forecasting model using FY-4A (FengYun-4A) satellite data and  Global Precipitation Measurement (GPM) IMERG (Integrated Multi-satellite Retrievals). Their model achieved a CC of approximately 0.58 and a CSI of 0.46 for a 0.5-hour lead time. In a separate study, \cite{ma2022improvement} employed a multiscale network that yielded a CC of 0.50 and a CSI of 0.24 against the NCEP Stage-IV analysis. The PrecipGradeNet model proposed by ~\cite{zhang2022precipgradenet} also demonstrated strong skill, with a CSI of 0.4 when validated against the IMERG Early Run. The higher-quality IMERG Final Run (FR), a multi-source benchmark, was not used as a real-time precipitation estimation due to its 3.5-month latency, which precludes real-time application.
% svm
% \cite{sehad2017novel}
% rf
% \cite{kuhnlein2014improving}

Thus, a clear gap remains: the need for a method that leverages the high fidelity of the IMERG FR benchmark without compromising the low latency required for real-time applications from geostationary orbit. To address this challenge, we introduce Huayu, a novel machine learning model for real-time precipitation estimation using infrared data from FY-4B Advanced Geostationary Radiation Imager (AGRI). By enhancing the GeoAttX framework~\citep{song5168620geoattx}, Huayu more effectively decodes the complex relationship between satellite infrared radiance and precipitation by training on the IMERG FR product. The development of Huayu focuses on two key advances: delivering improved estimation accuracy and achieving a substantial reduction in latency, providing a critical step towards operational, high-resolution, real-time precipitation monitoring from geostationary orbit. With regular dataset updates, Huayu will support policymakers and forecasters in monitoring weather events and responding rapidly to evolving hydrometeorological conditions. Further research will explore its applicability to specific scenarios and its potential for global operational deployment.

% The capacity of this precipitation to provide early warnings for floods and droughts is essential for disaster response, agricultural planning, and water management. 

% Precipitation is a critical weather parameter with profound scientific and societal importance. Accurate and real-time data are essential for disaster response, agricultural planning, and water resource management, enabling early warnings for floods and droughts. However, the high variability and intensity of rainfall make its accurate, large-scale estimation particularly challenging, as illustrated in Fig.~\ref{fig:introduction}.
% Huayu enhances both spatial and temporal resolution, achieving improvements from 0.1° and 30 minutes to 0.05° and 15 minutes, and shortens the latency from 3.5 months to 15 minutes. 
% Experimental results demonstrate that Huayu achieves CSI of 0.89 and CC of 0.7 through gridded validation, outperforming the above existing methods. More impressive independent third-party multi-source station validation results demonstrate that Huayu has better performance, with around a 3.74\% gain in CSI over IMERG ER, and stunning results that Huayu is advanced than IMERG FR (CSI: 0.693 vs. 0.670). IMERG FR is widely considered as the real precipitation in other works~\citep{zhang2022precipgradenet,zhu2024pecam}, and Huayu only costs one ten-thousandth latency (less than 15 minutes vs. 4 months) of it.

%% file: content/sections/4.experiments.tex
% \section{Experimental results and discussions}
\section{Results and discussion}
We evaluated the performance of Huayu alongside multiple mainstream precipitation products against rain gauge observations using a suite of key metrics. These included standard measures--Pearson Correlation Coefficient (CC), Root Mean Square Error (RMSE), Probability of Detection (POD), False Alarm Ratio (FAR), Accuracy (ACC), and Critical Success Index (CSI)--to assess overall and categorical skill. Among these, CSI is a key indicator of overall model performance as it balances the detection of hits (POD) against misses and false alarms (FAR). 
For a direct, pixel-scale comparison of precipitation intensities, we also calculated the Area Under the Receiver Operating Characteristic Curve (AUROC) and the Coefficient of Determination ($\rm R^2$) between Huayu and IMERG FR. 
To focus specifically on rainy areas, we employed a masked Pearson Correlation Coefficient ($\rm{CC_{rain}}$). This metric was computed by correlating Huayu and IMERG FR data exclusively at pixels where IMERG FR registered rainfall.

% defined as:
% \begin{equation}
% \label{eq:ccrain}
% \begin{aligned}
% M&=\{P_{i,j}...\in{P}|P_{i,j}= 0\},\\
%     {\rm mask}(X)&=\{X_{i,j}...\in{X}|X_{i,j}\not\in M\},\\
%     {\rm CC_{rain}}&={\rm CC}({\rm mask}(P_t^1),{\rm mask}(Y_t^1)).
% \end{aligned}
% \end{equation}
% Here, $M$ denotes the zero-value set derived from the ground truth precipitation data $P_t^1$, where each element of $P_t^1$ is represented as $P_{i,j}$ (with indices $i, j, \dots$ corresponding to the dataset dimensions, i.e., row and column indices for 2D spatial data). The function $\text{mask}(X)$ extracts a subset of elements from dataset $X$ (e.g., the predicted precipitation $Y_t^1$) by excluding those corresponding to zero-value positions in $P_t^1$. Specifically, only the elements in $X$ associated with non-zero positions in $P_t^1$ are retained. The $\text{CC}_{\text{rain}}$ metric is then computed as the Pearson Correlation Coefficient (CC) between the two masked datasets: $\text{mask}(P_t^1)$, representing the non-zero elements of the ground truth data, and $\text{mask}(Y_t^1)$ representing the corresponding elements of the predicted data. This approach ensures that zero-precipitation values, which are not the focus of accuracy assessment, are excluded from the correlation analysis.
\begin{figure}[!htbp]
\centering
\centerline{\includegraphics[width=1\textwidth]{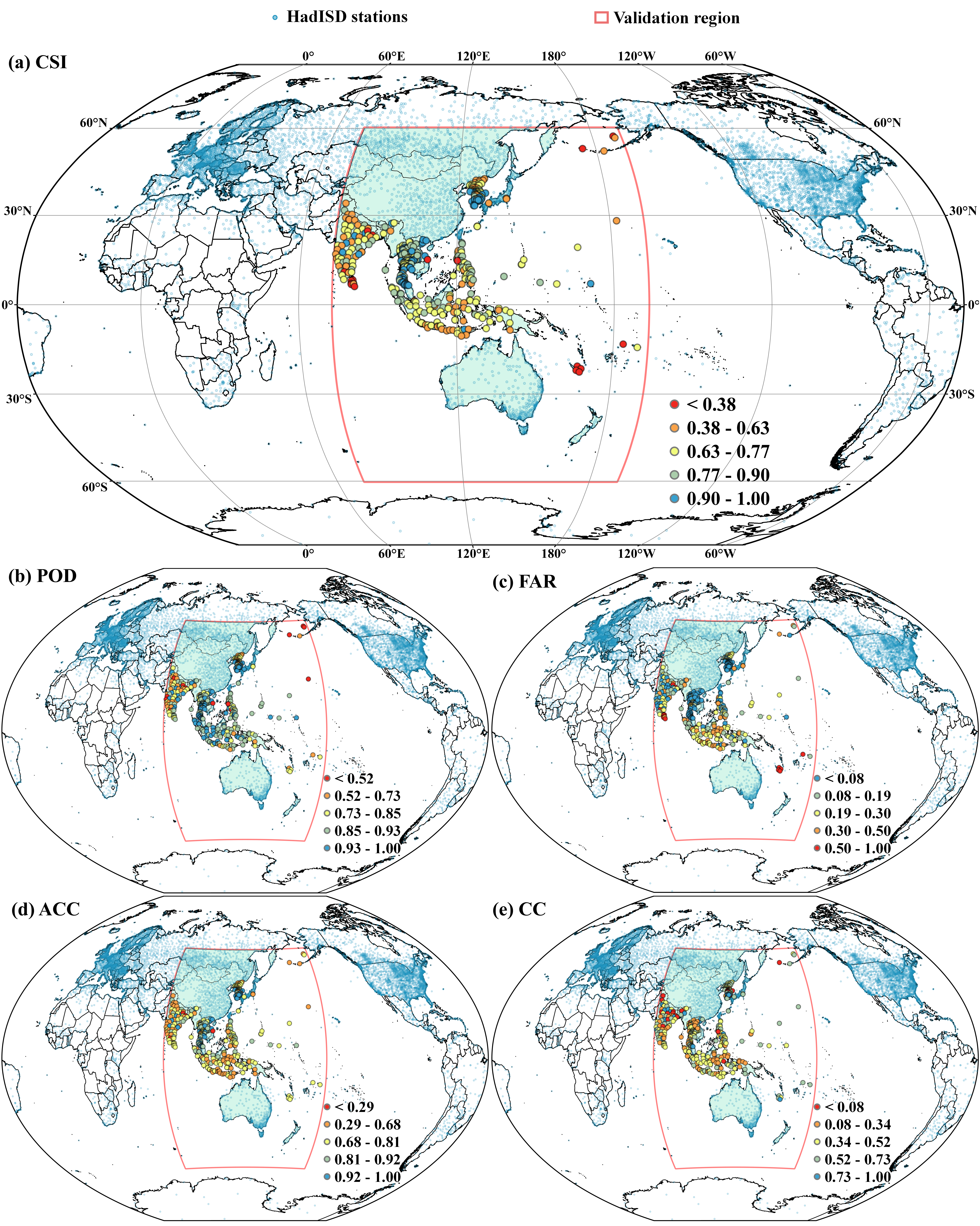}}
\caption{Distribution of HadISD stations and evaluation metrics for Huayu precipitation estimates:  (a) Critical Success Index (CSI), (b) Probability of Detection (POD), (c) False Alarm Ratio (FAR), (d) Accuracy (ACC), and (e) Pearson Correlation Coefficient (CC). Validation was conducted using data from 444 HadISD stations during July-December 2022. The analysis domain (red region) is centered at 133°E, corresponding to the nominal field for the FY-4B satellite. Note that the domain was repositioned to 105°E beginning 31 January 2024 (see Appendix Fig.~\ref{fig:region}). A total of 9,961 stations (shown as small blue dots) were excluded from analysis for being outside the study area or lacking valid observational data. All five evaluated metrics demonstrate consistently strong and coherent spatial performance.}

\label{fig:hadisd}
\end{figure}
\subsection{Validation against rain gauge observations}
\input{content/table/stations_metric}
We performed a continental-scale validation using the station-based HadISD dataset (version 3.4.3.2025f) \citep{dunn2014pairwise,dunn2012hadisd,smith2011integrated,dunn2016expanding} from 1 July to 31 December 2022. Due to the absence of hourly records, the analysis relied on three-hourly precipitation records. Of the initial 2,494 stations in the study area, 444 satisfied the data completeness criteria (Fig.~\ref{fig:hadisd}), yielding 55,087 valid records for analysis, with 65\% exhibiting rainfall.
The evaluation encompasses multiple operational global precipitation products: the FY-4B QPE (FngYun-4B Quantitative Precipitation Estimate, 15-min intermediate version), the NASA Global Precipitation Measurement (GPM) mission IMERG suite (ER, Late Run, FR), PERSIANN (Precipitation Estimation from Remotely Sensed Information using Artificial Neural Networks) and its variant PERSIANN-CCS, the GSMaP\_NOW (JAXA's Global Rainfall Map Realtime version), and the CMORPH (NOAA's Climate Prediction Center Morphing technique). The details about the above datasets are provided in Appendix~\ref{appsec:data}. The GeoAttX-based products under assessment include GeoAttX\_P, Huayu at 0.1° resolution (without downscaling), and Huayu at 0.05° resolution (implementing the downscaling strategy described in Sec.~\ref{sec:nrt}). Comprehensive evaluation results are presented in Tab.~\ref{tab:globe}.

Among all datasets validated against station gauges, Huayu demonstrated the best overall performance, leading in four key metrics: FAR, ACC, CSI, and CC. Although Huayu's Probability of Detection (POD: 0.874) was slightly lower than that of GeoAttX\_P (0.961), the latter achieved this high POD by frequently forecasting rain, resulting in the highest False Alarm Ratio (FAR: 0.353). Conversely, PERSIANN-CCS achieved a low FAR (0.186) but with the lowest POD (0.500), resulting in poor scores on comprehensive metrics like CSI (0.448) and CC (0.262). The standard PERSIANN model outperformed PERSIANN-CCS but still significantly trailed Huayu (CSI: 0.533 vs. 0.693). Meanwhile, the multi-source corrected CMORPH product--which has the latency of 3-4 months--delivered relatively strong performance, with the lowest overall FAR (0.176) and a CSI comparable to that of IMERGE FR (both around 0.6). Further validation details from rain gauge stations are provided in Appendix~\ref{appsec:shanghai} of the Supplementary Materials.

Huayu's superior performance compared to its training benchmark, IMERG FR, can be attributed to several factors. Firstly, its higher native resolution (0.05° vs. 0.1°) allows for a more precise match to station locations. Secondly, and more importantly, the machine learning process itself appears to act as a noise filter. While IMERG FR contains errors uncorrelated with true precipitation, Huayu, guided by the FY-4B imagery, cannot establish a physical link to these erroneous signals. Consequently, it learns to ignore this ``clutter'' and focuses on the most robust predictive features, effectively refining the original product. Finally, Huayu benefits from the unique observational perspective of the FY-4B satellite, which provides optimal coverage over Asia and Australia from its position at 133°E, a vantage point not fully leveraged by other global datasets~\citep{joyce2004cmorph,tan2019imerg,kubota2020global}. A detailed theoretical discussion is provided in the Supplementary material, Appendix~\ref{appsec:formula}.

\subsection{Evaluation across precipitation intensities at the pixel scale}

\begin{figure}[!htbp]
\centering
\centerline{\includegraphics[width=0.98\textwidth]{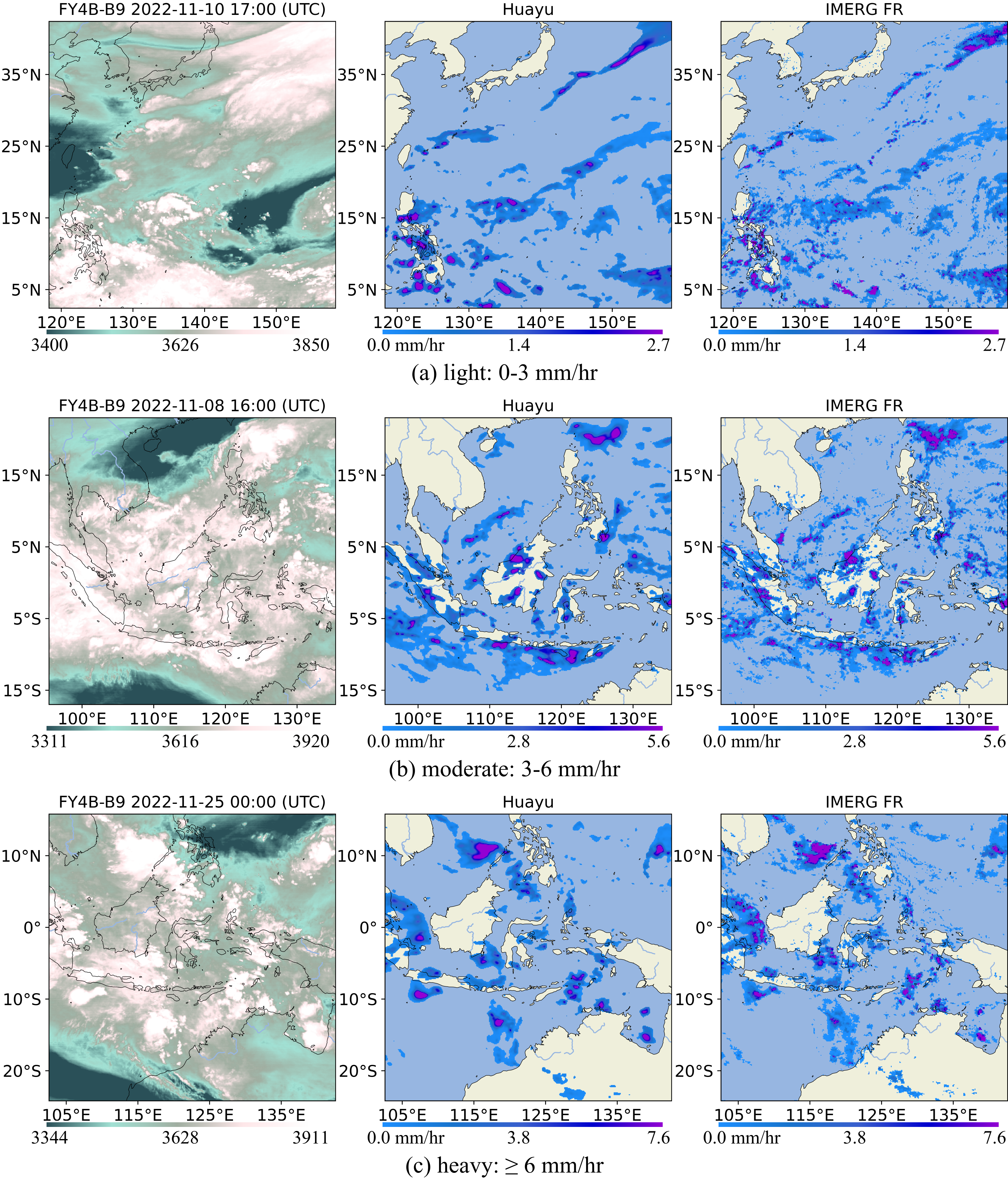}}
\caption{Visualization results for three validation samples across different precipitation intensity intervals (defined by 95th percentile maximum values: light $[0,3)$, moderate $[3,6)$, and heavy $\geq6$ mm/hr, respectively). ``FY4B-B9'' denotes the Band 9 data from AGRI instrument aboard the FengYun-4B satellite. ``Huayu'' represents the precipitation retrieved from the corresponding FY-4B AGRI bands 9-15, and ``IMERG FR'' serves as the benchmark reference (ground truth) in this comparison.}
\label{fig:visual}
\end{figure}
The validation analysis categorizes precipitation into three intensity levels based on the 95th percentile maximum rate: light, $[0,3)$; moderate, $[3,6)$; and heavy $\geq6$ mm/hr. We analyzed their characteristics in conjunction with the precipitation intensity. 
\input{content/table/metrics}
As illustrated in Tab.~\ref{tab:metrics}, the moderate precipitation interval ([3, 6) mm/hr) forms the dominant category in the validation set, comprising 495 samples (68.0\% of the total 728 samples). The model demonstrates progressively better performance at higher precipitation intensities, evidenced by increasing values of $\rm R^2$, CC, and $\rm CC_{rain}$. This correlation-based improvement pattern can be explained by the ability of these metrics to evaluate both spatial distribution and prediction accuracy simultaneously. It should be noted that RMSE, being inherently sensitive to precipitation magnitude, may not serve as a reliable standalone performance indicator. Apart from these four metrics ($\rm R^2$, CC, $\rm CC_{rain}$, and RMSE), the remaining five evaluation metrics showed no significant variations across intensity intervals. Consequently, the model exhibits lower predictive accuracy for light precipitation events compared to heavy precipitation scenarios.  

% POD and FAR represent two key evaluation metrics for precipitation inversion. Specifically, POD quantifies the precision of detection, whereas FAR indicates the reliability of the results. A higher POD value is preferable, as it reflects a better detection capability, while a lower FAR value is significantly more desirable, as it signifies fewer false alarms. Both metrics are bounded between 0 and 1. As demonstrated in Tab.~\ref{tab:metrics}, the POD and FAR of our results are satisfactory when compared against the ground truth data from IMERG. Furthermore, the CSI, which reaches a value of up to 0.89, serves as a comprehensive evaluation metric that underscores the high consistency between our inversion results and the target conditions. This metric is crucial for assessing both the reliability and practical applicability of our precipitation remote sensing inversion model.

Visualizations of model performance are provided in Fig.~\ref{fig:visual}. Each case displays the input FY-4B/AGRI 9th band channel (left), the precipitation estimated by our Huayu model (middle), and the target IMERG data for evaluation (right). The timestamps indicate the start of the imagery, with values representing the subsequent half-hour precipitation rate (mm/hr). The results show that Huayu successfully captures the spatial structure of the rainy areas, corroborating its strong performance in the POD, FAR, and CSI metrics reported in Tab.~\ref{tab:metrics}. Additional evaluations under typhoon-induced precipitation events are provided in Supplementary Appendix~\ref{appsec:typhoon}. In the context of climate change, which is amplifying the frequency and intensity of extreme precipitation events, accurate and real-time precipitation monitoring systems such as Huayu are increasingly critical for strengthening disaster early warning capabilities and enhancing the effectiveness of risk prevention and mitigation efforts.

% As shown in Fig.~\ref{fig:visual}, three precipitation intensity levels correspond to the metrics detailed in Tab.~\ref{tab:metrics}. The input data displayed in the left column represents the 9th band of AGRI on FY-4B, which serves as one of the initial field channels. The middle column presents the results generated by our proposed Huayu model, while the right column shows the target IMERG data used for performance evaluation. The time labels in the left column indicate the start time of the imagery, with the values in both the results and ground truth representing half-hour cumulative precipitation from the input start time. Case (a) exemplifies low precipitation intensities. As evident from the figure, the results successfully capture the contour of the rainy areas, indirectly confirming the model's superior performance in terms of POD, FAR, and CSI metrics as presented in Tab.~\ref{tab:metrics}. 

%% file: content/table/stations_metric.tex
\begin{table*}[t]
\scriptsize
\setlength{\extrarowheight}{2pt}
\centering
\caption{Performance evaluation of precipitation products using three-hourly station records (mm/hr). Six metrics are reported: Probability of Detection (POD),  False Alarm Ratio (FAR), Accuracy (ACC), Critical Success Index (CSI), CC (Pearson Correlation Coefficient), and Root Mean Square Error (RMSE), with arrows indicating the direction of improvement. Huayu (0.05°) demonstrates superior performance in key metrics (ACC, CSI, CC, and RMSE) against benchmark products, as highlighted in bold. The validation dataset (55,087 station records) is characterized by precipitation rate (35\% zeros, 40.63\% in $(0,1]$ mm/hr, 12.88\% in $(1,3]$ mm/hr, 4.59\% in $(3,5]$ mm/hr, and greater than 5 mm/hr is 6.53\%). The elevated rainfall intensities in these regions inflate RMSE values, making CSI and CC more robust metrics for comparison.}
% 0,3,9,15
% 19480	35.36%
% 22384	40.63%
% 7095	12.88%
% 2529	4.59%
% 3599	6.53%
\label{tab:globe} % 调整标签以区分原表格，避免引用冲突
\begin{tabular}{
p{2.2cm}<{\centering}  % 产品名称列（原表头）
p{1.6cm}<{\centering}
p{1.1cm}<{\centering}
p{0.7cm}<{\centering}
p{0.7cm}<{\centering}
p{0.7cm}<{\centering}
p{0.6cm}<{\centering}
p{0.4cm}<{\centering}
p{1.1cm}<{\centering}
}
\toprule
\tabincell{c}{\textbf{Product}}
&\tabincell{c}{\textbf{Resolution}}
&\tabincell{c}{\textbf{Latency}}
&\tabincell{c}{\textbf{POD}$\uparrow$}
&\tabincell{c}{\textbf{FAR}$\downarrow$}
&\tabincell{c}{\textbf{ACC}$\uparrow$}
&\tabincell{c}{\textbf{CSI}$\uparrow$}
&\tabincell{c}{\textbf{CC}$\uparrow$}
&\tabincell{c}{\textbf{RMSE}$\downarrow$\\(mm/hr)}
\\
\midrule
\tabincell{c}{FY-4B QPE} & \tabincell{c}{4km/15min} & \tabincell{c}{real-time} & \tabincell{c}{$0.518$} & \tabincell{c}{$0.186$} & \tabincell{c}{$0.611$} & \tabincell{c}{$0.463$} & \tabincell{c}{$0.258$} & \tabincell{c}{$9.230$} \\
\tabincell{c}{PERSIANN-CCS} & \tabincell{c}{0.04°/1hr} & \tabincell{c}{real-time} & \tabincell{c}{$0.500$} & \tabincell{c}{$0.186$} & \tabincell{c}{$0.603$} & \tabincell{c}{$0.448$} & \tabincell{c}{$0.262$} & \tabincell{c}{$3.335$} \\
\tabincell{c}{PERSIANN} & \tabincell{c}{0.25°/1hr} & \tabincell{c}{2day} & \tabincell{c}{$0.618$} & \tabincell{c}{$0.206$} & \tabincell{c}{$0.650$} & \tabincell{c}{$0.533$} & \tabincell{c}{$0.279$} & \tabincell{c}{$2.894$} \\
\tabincell{c}{GSMaP\_NOW} & \tabincell{c}{0.1°/1hr} & \tabincell{c}{real-time} & \tabincell{c}{$0.675$} & \tabincell{c}{$0.240$} & \tabincell{c}{$0.652$} & \tabincell{c}{$0.556$} & \tabincell{c}{$0.164$} & \tabincell{c}{$3.667$} \\
\tabincell{c}{CMORPH} & \tabincell{c}{8km/30min} & \tabincell{c}{3-4month} & \tabincell{c}{$0.699$} & \tabincell{c}{$\mathbf{0.176}$} & \tabincell{c}{$0.709$} & \tabincell{c}{$0.608$} & \tabincell{c}{$0.435$} & \tabincell{c}{$2.790$} \\
\tabincell{c}{GeoAttX\_P} & \tabincell{c}{0.05°/15min} & \tabincell{c}{real-time} & \tabincell{c}{$\mathbf{0.961}$} & \tabincell{c}{$0.353$} & \tabincell{c}{$0.636$} & \tabincell{c}{$0.630$} & \tabincell{c}{$0.237$} & \tabincell{c}{$4.436$} \\
\tabincell{c}{IMERG ER} & \tabincell{c}{0.1°/30min} & \tabincell{c}{4hr} & \tabincell{c}{$0.870$} & \tabincell{c}{$0.258$} & \tabincell{c}{$0.721$} & \tabincell{c}{$0.668$} & \tabincell{c}{$0.382$} & \tabincell{c}{$2.780$} \\
\tabincell{c}{IMERG LR} & \tabincell{c}{0.1°/30min} & \tabincell{c}{12hr} & \tabincell{c}{$0.858$} & \tabincell{c}{$0.249$} & \tabincell{c}{$0.725$} & \tabincell{c}{$0.668$} & \tabincell{c}{$0.426$} & \tabincell{c}{$2.708$} \\
\tabincell{c}{IMERG FR} & \tabincell{c}{0.1°/30min} & \tabincell{c}{3.5month} & \tabincell{c}{$0.863$} & \tabincell{c}{$0.250$} & \tabincell{c}{$0.726$} & \tabincell{c}{$0.670$} & \tabincell{c}{$0.441$} & \tabincell{c}{$2.666$} \\
\midrule
\tabincell{c}{\textbf{Huayu}} & \tabincell{c}{0.1°/15min\\0.05°/15min} & \tabincell{c}{real-time} & \tabincell{c}{$0.872$\\$0.874$} & \tabincell{c}{$0.230$\\$0.230$} & \tabincell{c}{$0.749$\\$\mathbf{0.750}$} & \tabincell{c}{$0.692$\\$\mathbf{0.693}$} & \tabincell{c}{$0.445$\\$\mathbf{0.452}$} & \tabincell{c}{$2.636$\\$\mathbf{2.622}$} \\
% \tabincell{c}{\textbf{Huayu}} & \tabincell{c}{} & \tabincell{c}{real-time} & \tabincell{c}{} & \tabincell{c}{} & \tabincell{c}{} & \tabincell{c}{$\mathbf{0.693}$} & \tabincell{c}{$\mathbf{0.452}$} & \tabincell{c}{$\mathbf{2.622}$} \\

\bottomrule
\end{tabular}
\end{table*}

%% file: content/table/metrics.tex
\begin{table*}[!t]
% \tiny
\scriptsize
% \fontsize{6}{6}
\setlength{\extrarowheight}{2pt}
  \centering
  \caption{Performance metrics across precipitation intensity intervals. Nine evaluation metrics are computed on the validation samples (n = 728) stratified by the 95th percentile maximum value of precipitation intensity (light $[0,3)$, moderate $[3,6)$, and heavy $\geq6$ mm/hr) (see Supplementary Appendix~\ref{appsec:shanghai}). Lower values indicate better performance for FAR and RMSE, while higher values are desirable for the remaining seven metrics.}
  \label{tab:metrics}
  % \resizebox{\textwidth}{!}{
  \begin{tabular}
  {p{1.3cm}<{\centering}
  % p{2.5cm}<{\centering}
  p{0.7cm}<{\centering}
  p{0.4cm}<{\centering}
  p{0.4cm}<{\centering}
  p{0.8cm}<{\centering}
  p{1cm}<{\centering}
  p{0.6cm}<{\centering}
  p{0.6cm}<{\centering}
  p{0.6cm}<{\centering}
  p{0.5cm}<{\centering}
  p{0.9cm}<{\centering}
  }
 \toprule
 \tabincell{c}{{\fontsize{8}{0}\textbf{Intensities}}\\{\fontsize{6}{0}\selectfont(mm/hr)}}
&\tabincell{c}{\fontsize{7}{8}\textbf{Sample}\\\textbf{Size}}
&\tabincell{c}{\fontsize{7}{8}$\mathbf{R^2}$}
&\tabincell{c}{\fontsize{7}{8}$\mathbf{CC}$}
&\tabincell{c}{\fontsize{7}{8}$\mathbf{CC_{rain}}$}
&\tabincell{c}{\fontsize{7}{8}$\mathbf{AUROC}$}
&\tabincell{c}{\fontsize{7}{8}$\mathbf{POD}$}
&\tabincell{c}{\fontsize{7}{8}$\mathbf{FAR}$}
&\tabincell{c}{\fontsize{7}{8}$\mathbf{ACC}$}
&\tabincell{c}{\fontsize{7}{8}$\mathbf{CSI}$}
&\tabincell{c}{{\fontsize{7}{8}$\mathbf{RMSE}$}\\{\fontsize{6}{0}\selectfont(mm/hr)}}
\\
\midrule[0.75pt]

\tabincell{c}{$\mathbf{[0,3)}$}
&\tabincell{c}{114}
&\tabincell{c}{0.35}
&\tabincell{c}{0.61}
&\tabincell{c}{0.54}
&\tabincell{c}{0.94}
&\tabincell{c}{0.96}
&\tabincell{c}{0.05}
&\tabincell{c}{0.92}
&\tabincell{c}{0.91}
&\tabincell{c}{0.33}
\\
\tabincell{c}{$\mathbf{[3,6)}$}
&\tabincell{c}{495}
&\tabincell{c}{0.49}
&\tabincell{c}{0.71}
&\tabincell{c}{0.65}
&\tabincell{c}{0.94}
&\tabincell{c}{0.94}
&\tabincell{c}{0.07}
&\tabincell{c}{0.90}
&\tabincell{c}{0.88}
&\tabincell{c}{0.70}
\\
\tabincell{c}{$\mathbf{\geq 6}$}
&\tabincell{c}{119}
&\tabincell{c}{0.54}
&\tabincell{c}{0.74}
&\tabincell{c}{0.69}
&\tabincell{c}{0.95}
&\tabincell{c}{0.94}
&\tabincell{c}{0.06}
&\tabincell{c}{0.91}
&\tabincell{c}{0.89}
&\tabincell{c}{0.92}
\\
\midrule
\tabincell{c}{\textbf{Overall}}
&\tabincell{c}{728}
&\tabincell{c}{0.48}
&\tabincell{c}{0.70}
&\tabincell{c}{0.64}
&\tabincell{c}{0.94}
&\tabincell{c}{0.94}
&\tabincell{c}{0.07}
&\tabincell{c}{0.90}
&\tabincell{c}{0.89}
&\tabincell{c}{0.68}
\\
\bottomrule
\end{tabular}
% }
\end{table*}

%% file: content/sections/3.methods.tex
\begin{figure}[th!]
\centering
\centerline{\includegraphics[width=0.95\textwidth]{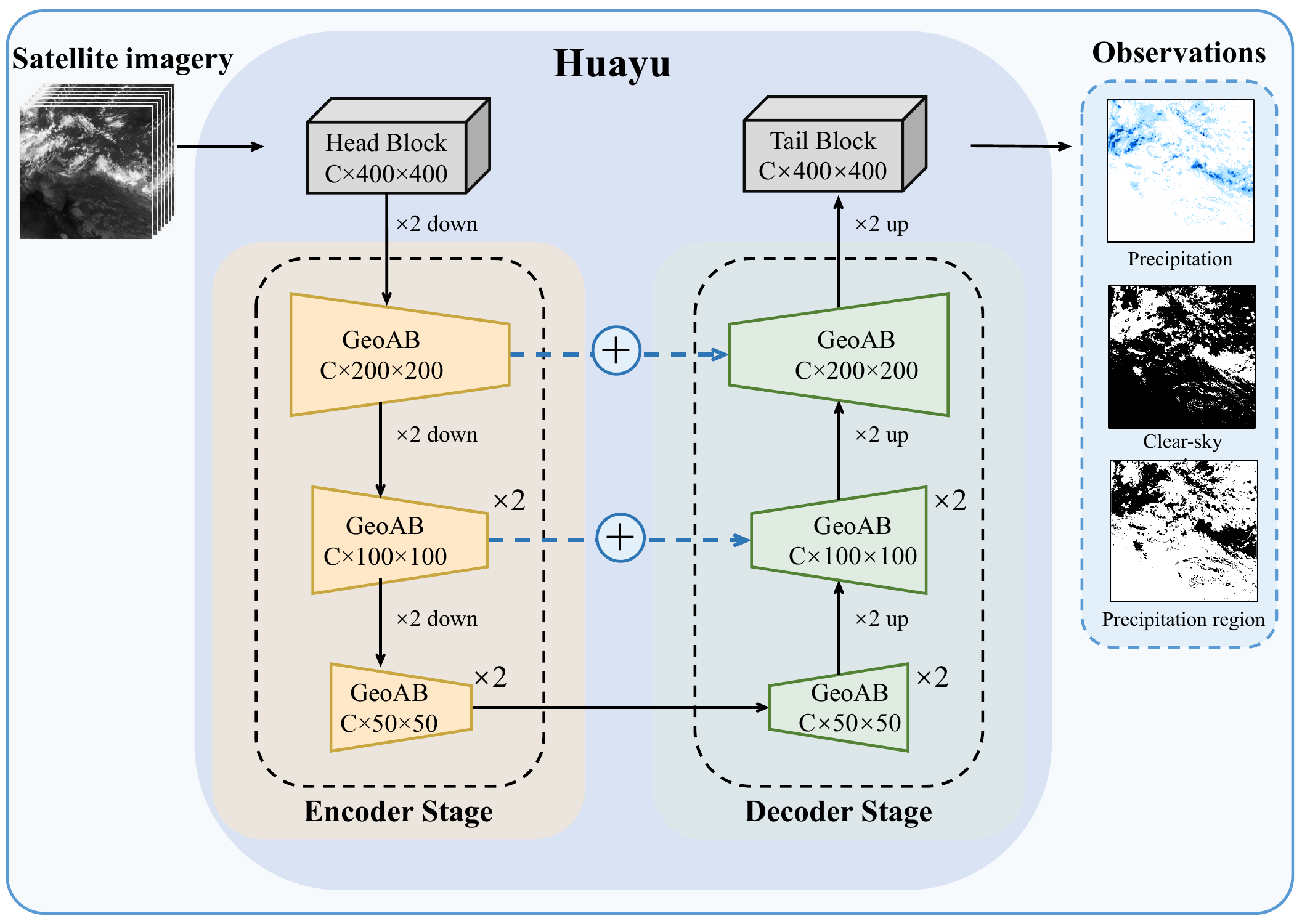}}
\caption{The structure of the proposed network Huayu. The head block, tail block, and GeoAB block are defined in the GeoAttX framework. There are five GeoABs in both the encoder and decoder stages.}
\label{fig:structure}
\end{figure}

\begin{figure}[th!]
\centering
\centerline{\includegraphics[width=0.7\textwidth]{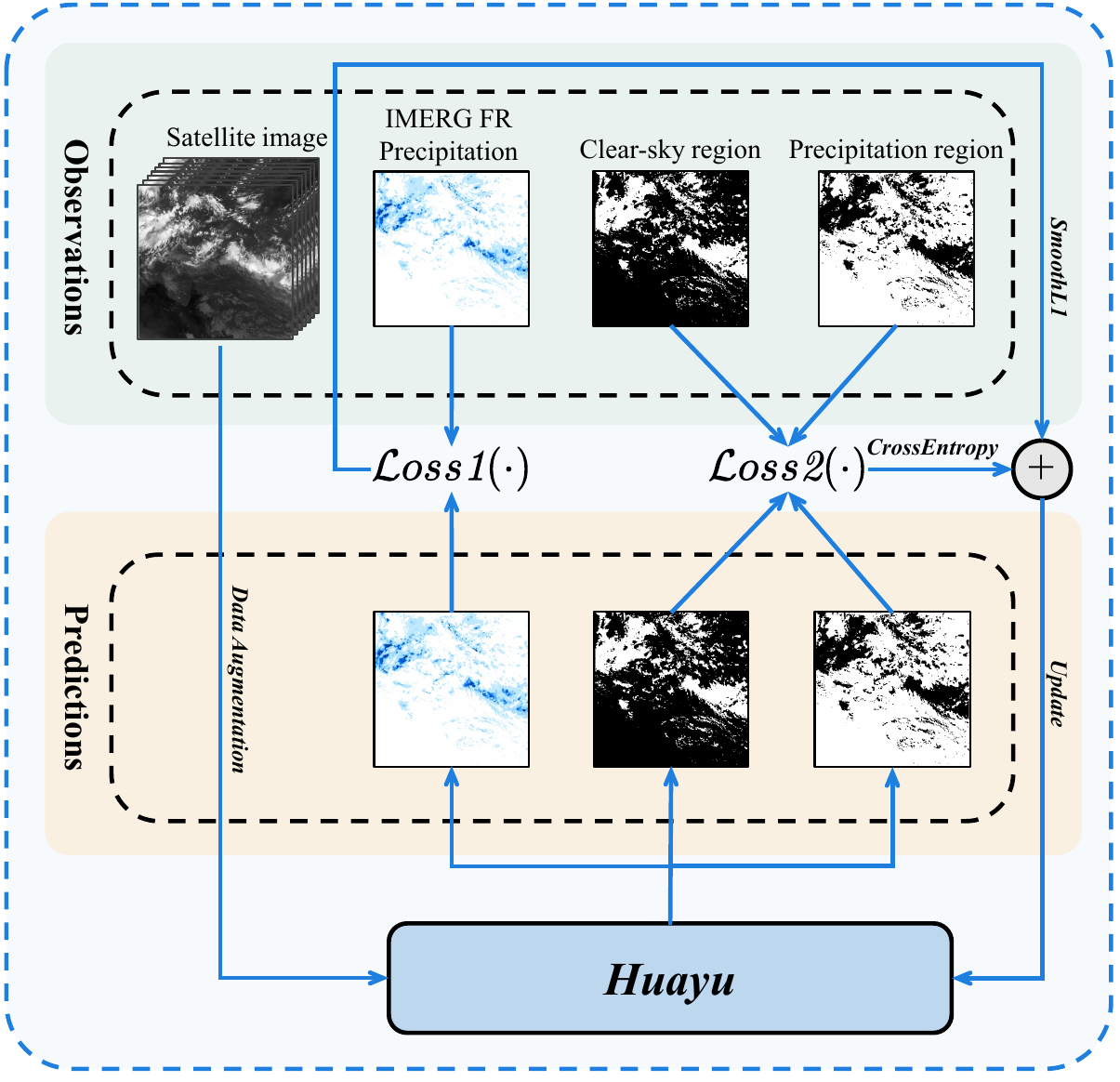}}
\caption{Dual-target loss for training, applied to randomly sampled FY-4B AGRI patches (7 bands) and corresponding IMERG data (precipitation rate, clear-sky, and precipitation region masks).}
\label{fig:doubleloss}
\end{figure}

% \section{Methodology}
\section{Methods}
% \subsection{Network structure}
Geographic Attention Block (GeoAB)~\citep{song5168620geoattx} was designed for precipitation forecasting by two steps: (1) GeoAttX\_I, which focuses on cloud image extrapolation, and (2) GeoAttX\_P, which is used for precipitation estimation. Based on GeoAB, this study proposes an advanced satellite-based precipitation estimation model named Huayu, 
% The latter model relies on Quantitative Precipitation Estimation (QPE) data derived from the AGRI sensor. However, QPE's applicability is constrained by its limited data sources and its origin as a FengYun-specific product, which results in restricted diversity of input information. Furthermore, GeoAttX\_P operates on a fixed grid structure, limiting its adaptability to data obtained from other satellites. Notably, IMERG is a more widely recognized and globally utilized satellite precipitation estimation dataset than QPE.
which was further improved through the incorporation of a ``U'' structure \citep{ronneberger2015u,chen2023fuxi} as illustrated in Fig.\ref{fig:structure}. The input satellite imagery is represented by $F_t$, and the output $Y_t$ is computed by Huayu according to the following equation:
\begin{equation}
\label{eq:main}
    Y_t={\rm Huayu}(F_t).
\end{equation}
More details about the setting of Huayu are provided in Appendix~\ref{appsec:setting}, and the implementation details are available in the source code.
\subsection{Dual-target loss}
Deep learning faces an inherent challenge in regression-based precipitation estimation: neural networks tend to produce values near zero rather than exact zeros, making it difficult to distinguish between light rain and clear skies. To overcome this limitation, Huayu employs a two-stage dual-target loss. It first performs binary classification to identify rainy regions, and subsequently generates precipitation values only for areas classified as rainy. Furthermore, the regression loss is computed exclusively over these rainy regions, thereby focusing the model's capacity on accurate precipitation estimation where it matters most. The dual-target loss function is defined as:
\begin{equation}
\label{eq:loss}
\begin{aligned}
    loss = {\rm SmoothL1}(P^1_t, Y^1_t) + {\rm CrossEntropy}(P^3_t, Y^2_t, Y^3_t).
\end{aligned}
\end{equation}
The dual-target loss combines two components: the ${\rm CrossEntropy}$ term identifies rainy regions, while the ${\rm SmoothL1}$ term regulates precipitation amounts, as shown in Fig.~\ref{fig:doubleloss}. The first channel (i.e., precipitation rate) of the target data \( P_t \) at time \( t \) is denoted as \( P^1_t \) and \( Y^1_t \) in the model output, and so on.

\begin{figure}[th!]
\centering
\centerline{\includegraphics[width=1\textwidth]{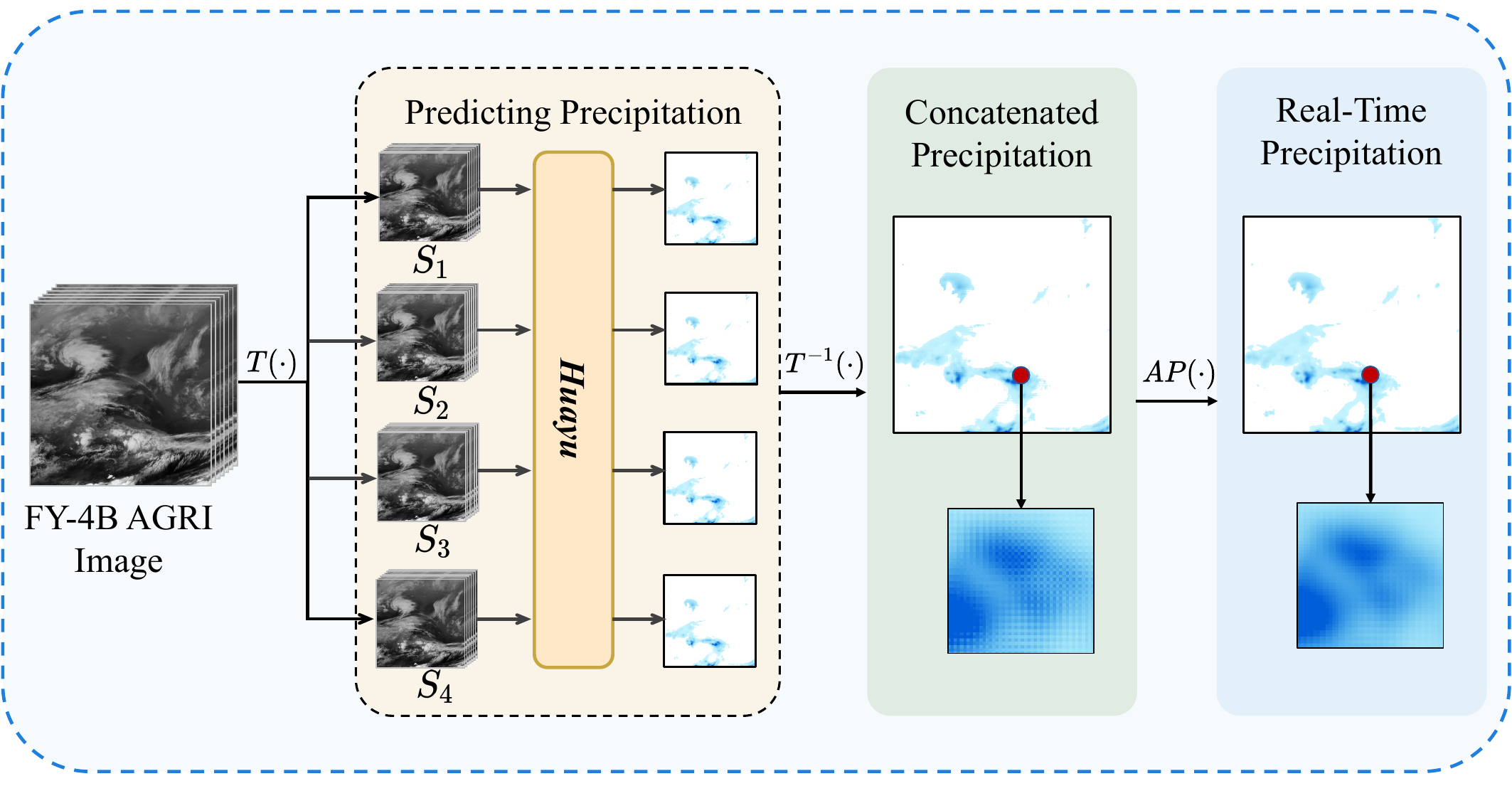}}
\caption{The process of Huayu real-time system from FY-4B. After the recalibration, the satellite image is clipped into four sub-pieces ($S_1$-$S_4$) by the operation $T(\cdot)$. Huayu will produce four corresponding precipitation for the next reversed $T(\cdot)$ (i.e., $T(\cdot)^{-1}$) to achieve a 0.05° total precipitation in 15 minutes. To optimize the grid effects, the operation $AP(\cdot)$ was adopted to produce the final real-time precipitation.}
\label{fig:nrt}
\end{figure}

\subsection{Huayu real-time system}
\label{sec:nrt}
To achieve real-time satellite precipitation estimation, it is essential to utilize the FY-4B AGRI data, which typically becomes available with a 15-minute latency, in time. After recalibration and clipping, AGRI data is subdivided into four images as illustrated in Fig.~\ref{fig:nrt} through the operation denoted by $T(\cdot)$, and each sub-image has a resolution of $0.1^\circ \times 0.1^\circ$. As formalized in Eq.~\ref{eq:split}, the transformation \( T(\cdot) \) decomposes an original image \( S \) into four non-overlapping sub-images \( S_1, S_2, S_3, S_4 \). This is achieved by defining two index sets - \( E \) for even indices and \( O \) for odd indices - within the range below 800, corresponding to the spatial dimensions of the input FY-4B imagery. The decomposition effectively performs a non-overlapping 2×2 block operation, where each sub-image is constructed by selecting pixels from specific positions within each block. Formally, this is expressed as: 
\begin{equation}
\begin{aligned}
T(S) &\;=\; S_1, \, S_2, \, S_3, \, S_4, \\[6pt]  % 增加垂直间距区分逻辑组
E    &\;=\; \{0, \, 2, \, \ldots, \, 798\}, O \;=\; \{1, \, 3, \, \ldots, \, 799\}, \\[6pt]  % 增加组间间距
S_1  &\;=\; \{ T_{i,j} = S_{i,j} \mid i \in E,\, j \in E \}, S_2 \;=\; \{ T_{i,j} = S_{i,j} \mid i \in O,\, j \in E \}, \\
S_3  &\;=\; \{ T_{i,j} = S_{i,j} \mid i \in E,\, j \in O \}, S_4 \;=\; \{ T_{i,j} = S_{i,j} \mid i \in O,\, j \in O \}.
\end{aligned}
\label{eq:split}
\end{equation}
Each sub-image is independently processed by Huayu to generate a corresponding sub-precipitation estimate at 0.1°×0.1° resolution, as shown in Fig.~\ref{fig:nrt}. These estimates are then merged via the inverse transformation \( T^{-1}(\cdot) \) into a unified precipitation field at 0.05°×0.05° resolution. To reduce grid artifacts introduced during the block-wise decomposition, an average pooling operation is applied during reconstruction. The merged precipitation field \(P_{i,j}\) is defined over an \(800 \times 800\) grid, with zero-based indices \(i,j \in \{0, 1, \dots, 799\}\). A \(3 \times 3\) average pooling is applied to interior pixels:
\[
\frac{1}{9} \sum_{m=i-1}^{i+1} \sum_{n=j-1}^{j+1} P_{m,n},
\]
which smooths high-frequency discontinuities while preserving spatial details. For edge pixels, the original values \(P_{i,j}\) are retained to keep the image size \(800 \times 800\). This strategy effectively suppresses grid effects, visible block boundaries caused by grayscale inconsistencies between adjacent blocks, without compromising image quality.

%% file: content/sections/6.information.tex
\section*{Acknowledgements}
This work is supported by National Natural Science Foundation of China under Grant No. W2412140, 42230505, and 42471079, and International Research Center of Big Data for Sustainable Development Goals under Grant No. CBAS2022GSP07. We thank all the datasets used in this work, including IMERG, FY-4B full disk imagery, FY-4B QPE, HadISD, PERSIANN, CMORPH, GSMaP, and the stations from the Shanghai Water Authority.

% \section*{Open Research}
\section*{Data availability}
The training dataset of \href{https://satellite.nsmc.org.cn/PortalSite/Data/DataView.aspx}{FengYun-4B geostationary meteorological satellite imagery and FY-4B QPE} are available from the National Satellite Meteorological Center. Geo-calibration procedures followed the official FY-4B guidelines, which are available on the official website. \href{https://gpm.nasa.gov/data/imerg}{IMERG data} were obtained from NASA. The \href{https://swj.sh.gov.cn/}{Shanghai station data} were provided by the Shanghai Water Authority (Shanghai Municipal Oceanic Bureau), and the globe station data \href{https://www.metoffice.gov.uk/hadobs/hadisd/}{HadISD} were provided by the Met Office. \href{http://chrsdata.eng.uci.edu/}{PERSIANN and PERSIANN-CCS} can be accessed by the official website. \href{http://sharaku.eorc.jaxa.jp/GSMaP_NOW/}{GAMaP\_NOW} is provided by JAXA, and \href{https://www.ncei.noaa.gov/products/climate-data-records/precipitation-cmorph}{CMORPH} data can be downloaded from the website of NOAA.

\section*{Code availability}
The source code for Huayu is accessible in \href{https://github.com/songzijiang/jacksung/blob/main/jacksung/ai/GeoAttX.py}{jacksung package}. For a quick start, please refer to \href{https://github.com/songzijiang/GeoAttX}{this repository}. The result of Huayu by using the utility package is total precipitation (mm) in 15 minutes, same as the GeoAttX\_P. For further inquiries regarding the utilization of the GeoAttX model series, interested parties may contact the authors via email.

% \section*{CRediT authorship contribution statement}
\section*{Author contributions}
\textbf{Zijiang Song}: Writing - original draft, Methodology. \textbf{Ting Liu}: Formal analysis, Visualization. \textbf{Lina Yuan}: Writing - review \& editing, Supervision. \textbf{Yuying Li}: Validation. \textbf{Ao Xu}: Visualization. \textbf{Xigang Sun}: Methodology. \textbf{Ye Li}: Writing – review \& editing, Funding acquisition. \textbf{Feng Lu}: Writing – review \& editing, Validation. \textbf{Min Liu}: Supervision, Resources, Funding acquisition. 

% \section*{Declaration of competing interest}
\section*{Competing interests}
The authors declare no competing interests.

%% file: content/sections/Supplementary.tex
\begin{appendices}
% \section{Section title of first appendix}\label{secA1}
\input{content/sections/2.data}

\section{More discussion about performance}
\label{appsec:formula}
\begin{figure}[thbp!]
\centering
\centerline{\includegraphics[width=0.6\textwidth]
{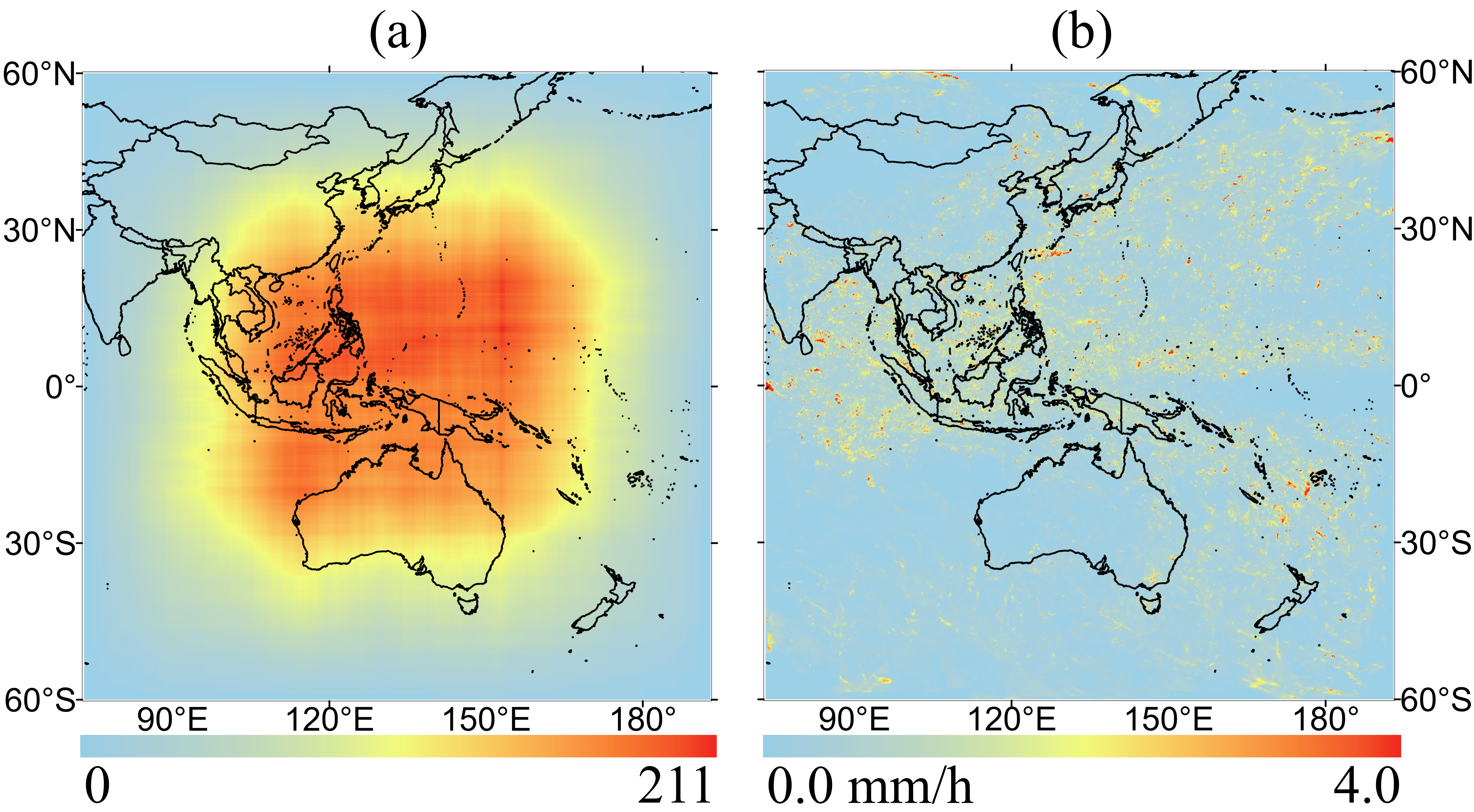}}
\caption{Error distribution on the validation dataset: (a) heatmap of sample density across pixels, and (b) pixel-wise root mean square error (RMSE).}
\label{fig:err}
\end{figure}
\begin{figure}[th!]
\centering
\centerline{\includegraphics[width=1\textwidth]{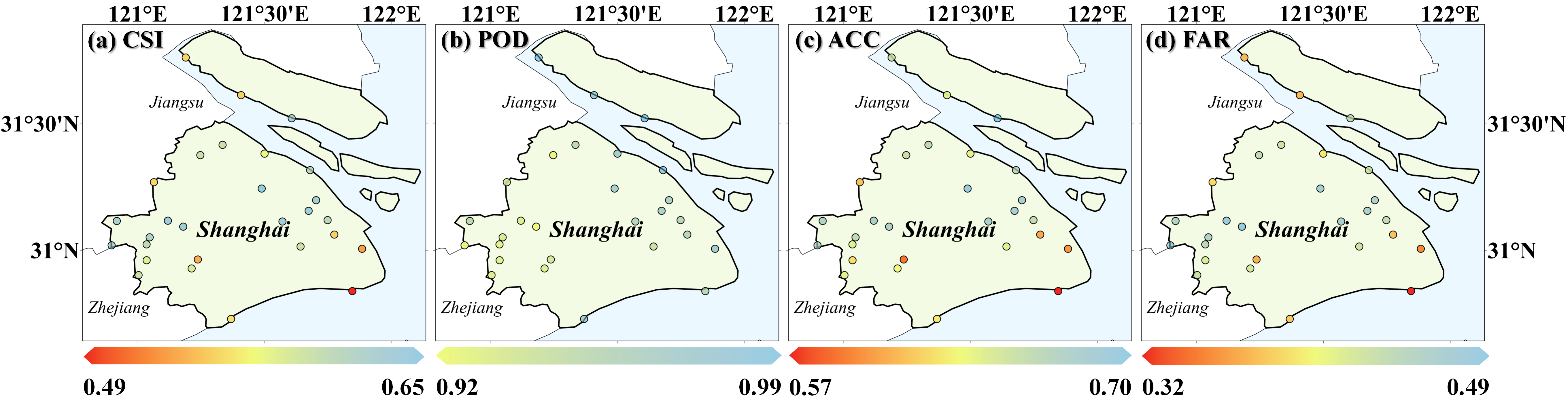}}
\caption{Four metrics (CSI, POD, ACC, and FAR) of Huayu on 28 stations in Shanghai.}
\label{fig:stations}
\end{figure}
Here are the detailed formulas for the speculation of why Huayu outperforms IMERG FR. The number of Huayu parameters is 241.78M recorded as $N$, each parameter is recorded as $P_i$ ($i=1\dots N$). IMERG FR stands by $I$, and the real precipitation is $O$. The errors between IMERG FR and the real precipitation are $\epsilon$, (i.e., $I=O+\epsilon$)~\citep{tan2016novel}. The relationship from FY-4B (recoreded as $FY$) to IMERG FR is recorded as $\mathcal{H}(\cdot)$, which is determined by $P$ ($P=\{P_i, i=1\dots N\}$). Then we have the formula:
\begin{equation}
\label{eq:discussion}
\begin{aligned}
\mathcal{H}(FY)&\rightarrow I.
\end{aligned}
\end{equation}
Assume the first $\omega$ parameters contribute to the $\epsilon$ and the rest contribute to the $O$. Then, we can get the following formula:
\begin{equation}
\label{eq:suberr}
\begin{aligned}
% \mathcal{H}_{(1,\omega)}(FY)\rightarrow \epsilon,\\
% \mathcal{H}_{(\omega,N)}(FY)\rightarrow O,\\
\mathcal{H}_{(1,\omega)}(FY)+\mathcal{H}_{(\omega,N)}(FY)&\rightarrow \epsilon+O.
\end{aligned}
\end{equation}
Assume the ideal relationship from FY-4B to real precipitation is $\mathcal{I}(\cdot)$ (i.e., $\mathcal{I}(FY)=O$). With the increasing of the training epochs, $\mathcal{H}_{(1,\omega)}(\cdot)$ will converge to zero, and $\mathcal{H}_{(\omega,N)}(\cdot)$ will coverge to $\mathcal{I}(\cdot)$, as Huayu cannot found the relation from $FY$ to $O$ leading to the parameters weight of $P_i$ ($i=1\dots \omega$) decline. Then the relationship of Huayu is updated as follows:
\begin{equation}
\label{eq:merge}
\begin{aligned}
\mathcal{H}(FY)\approx \mathcal{H}_{(\omega,N)}(FY)&\rightarrow \mathcal{I}(FY)=O\\
\mathcal{H}(FY)&\rightarrow O.
\end{aligned}
\end{equation}
Thus, our proposed Huayu could only learn the existing relationship between FY-4B and real precipitation, ignoring the errors inherently existing in the IMERG FR.
\input{content/table/shanghaigauge}
\begin{figure}[th!]
\centering
\centerline{\includegraphics[width=0.98\textwidth]{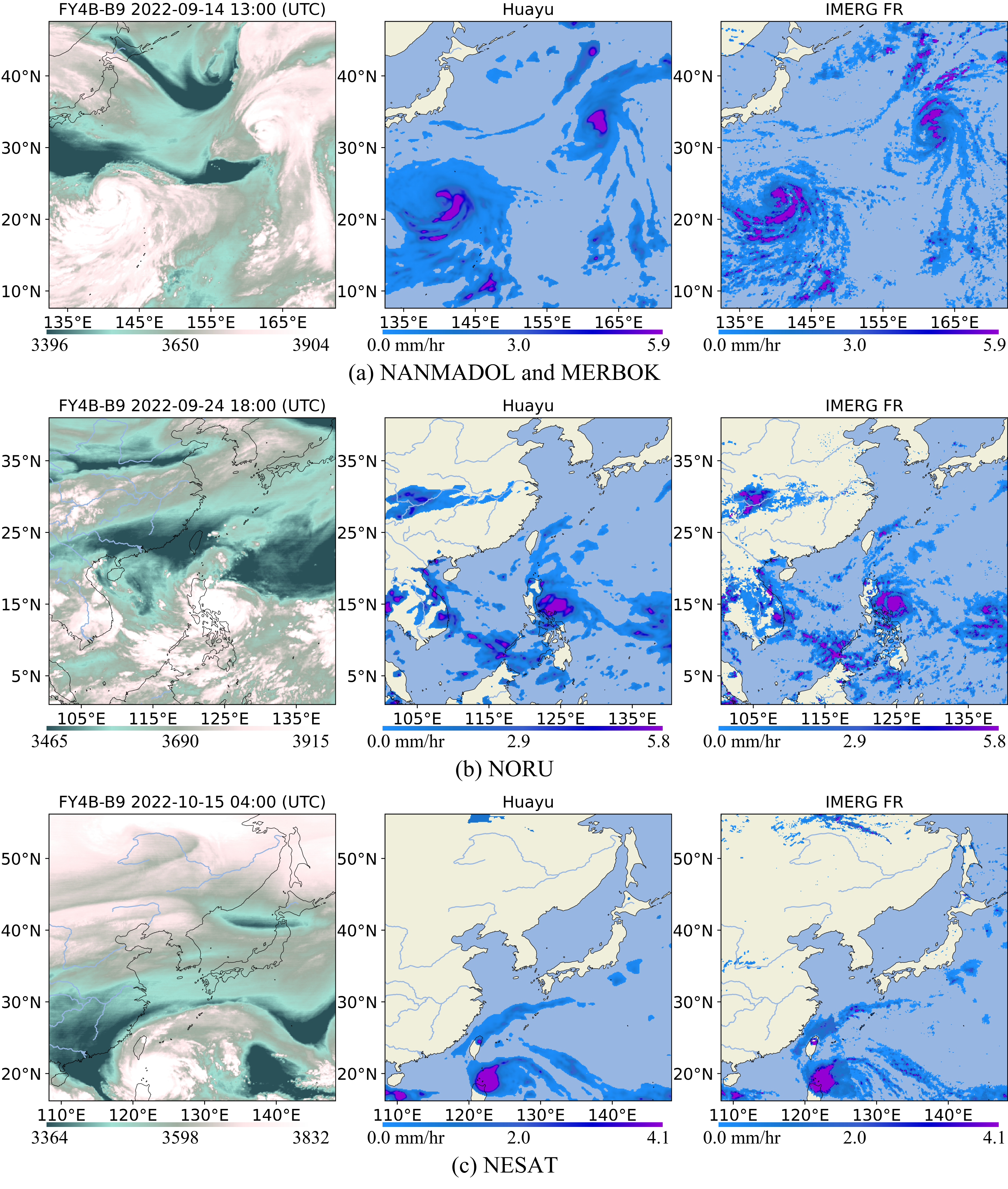}}
\caption{The visual precipitation estimation gridded comparison between Huayu and IMERG FR of four chosen typhoons in three cases: (d) 2214 NANMADOL (left) and 2213 MERBOK (right); (e) 2216 NORU; (f) 2220 NESAT, in the validation dataset in 2022.}
\label{fig:typhoon}
\end{figure}
\section{Extented Results}
\label{appsec:validation}
\subsection{RMSE of the validation dataset}
Figure~\ref{fig:err} shows the RMSE distribution for Huayu. As panels (a) and (b) indicate, higher precipitation occurs in the equatorial region (10°S-10°N) due to atmospheric circulation. This region also exhibits a higher RMSE, which can be attributed to two factors: on one hand, extreme precipitation values in this area present a known challenge for deep learning methods, making accurate prediction difficult; on the other hand, the overall high precipitation levels in the region itself also contribute to the elevated RMSE.
\subsection{Evaluation of the Gauge in Shanghai}
\label{appsec:shanghai}
We evaluated the generated precipitation against records from 28 rain gauges in Shanghai, sourced from the \href{https://swj.sh.gov.cn/}{Shanghai Water Authority (Shanghai Municipal Oceanic Bureau)}. Given that the period between May and August represents the concentrated rainfall season in Shanghai, measured data specifically from this interval (May-August 2025) were selected for analysis to more effectively assess the models’ performance during actual precipitation events. The evaluation outcomes across 28 individual stations with 289 precipitation cases, as presented in Tab.~\ref{tab:stations} and Fig.~\ref{fig:stations}, provided additional insights into the spatial distribution of model performance. As indicated in Tab.~\ref{tab:stations}, Huayu exhibited superior capability in detecting precipitation events, achieving a CSI score of up to 0.60, which is notably higher than ER's 0.57, even if Huayu has short latency. 

\begin{figure}[th]
\centering
\centerline{\includegraphics[width=1\textwidth]{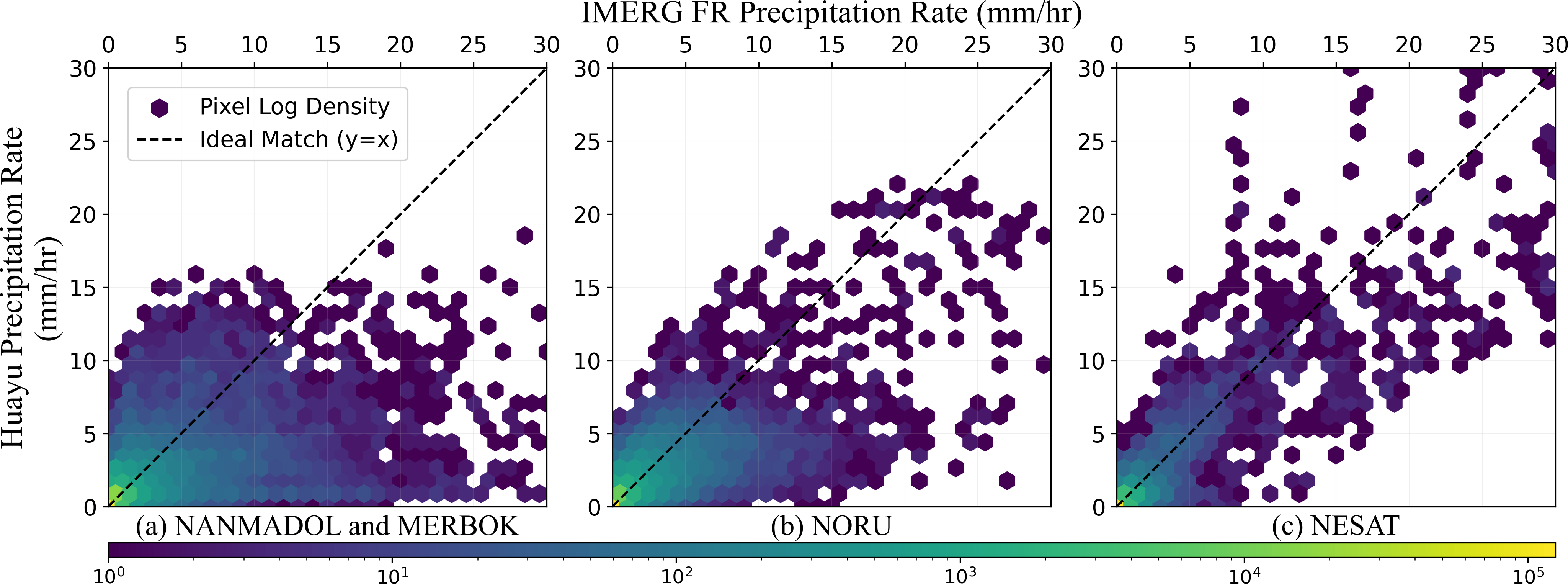}}
\caption{The pixel level distribution of four typhoons: NANMADOL, MERBOK, NORU, and NESAT, when comparing the Huayu and IMERG FR. The hexagon represents the pixel density of log, and the dashed line represents the ideal results.}
\label{fig:distribution}
\end{figure}

\subsection{Performance during Typhoon-induced precipitation events}
\label{appsec:typhoon}
To evaluate the model's performance under typhoon conditions, we present four distinct typhoons—NANMADOL, MERBOK, NORU, and NESAT—across three cases in Fig.~\ref{fig:typhoon}. Case (d) depicts two typhoons simultaneously: NANMADOL (left) and MERBOK (right). Huayu successfully reconstructs the general outline of typhoon precipitation patterns. However, as shown in Fig.~\ref{fig:distribution}, finer details are less accurately captured. While the model achieves a high correlation with ground truth data for typhoons NORU and NESAT, it exhibits a consistent tendency to underestimate precipitation rates across all cases. These results indicate that while Huayu captures the overall distribution of typhoon-related precipitation, its simulation of fine-grained features requires improvement. Future work will focus on enhancing the representation of detailed precipitation processes through optimized parameterization schemes or the integration of additional high-resolution observational data.

\section{Parameter settings}
\label{appsec:setting}
Huayu contains 241.78M parameters. To accommodate GPU memory constraints, the network uses 90 channels per feature map and 10 GeoAB blocks. The model was trained on data from 2023 and 2024. To reduce redundancy from consecutive, similar images, a 6-hour sampling interval was applied. For data augmentation, each original 2,400×2,400 FY-4B image was randomly cropped into 9 sub-images. We also excluded samples where precipitation covered less than 2\% of the area to address class imbalance. The final dataset consisted of 23,165 training and 728 validation samples. Training was conducted for 50 epochs using the AdamW optimizer~\citep{kingma2014adam,loshchilov2017decoupled} with an initial learning rate of 0.001, which was halved every 10 epochs. We used a batch size of 32 on four NVIDIA RTX 6000 Ada GPUs (48 GB memory each), completing the training in two days. Our implementation uses PyTorch with Python 3.12 on Linux.
\end{appendices}

%% file: content/sections/2.data.tex
\section{Related data}
\label{appsec:data}
Geostationary Meteorological Satellites (GMS), positioned at an altitude of 35,786 km above the equator, adhere to a geostationary orbit. This orbital configuration enables these satellites to maintain a stationary position relative to specific terrestrial regions by synchronizing their orbital period with Earth's rotational cycle. The presently operational geostationary meteorological satellites include FengYun-4A/B~\citep{yang2017introducing}, GOES-16/18~\citep{schmit2005introducing,schmit2017closer}, Himawari-9~\citep{bessho2016introduction}, and Meteosat-11/12~\citep{schmetz2002introduction}, whose parameters are summarized in Tab.~\ref{tab:satellite}. The Advanced Baseline Imager (ABI), Advanced Himawari Imager (AHI), and Flexible Combined Imager (FCI) are equipped with 16 spectral bands, providing a resolution of 2 km/10 min. In contrast, the spatial and temporal resolutions of the Advanced Geostationary Radiation Imager (AGRI) and Spinning Enhanced Visible and Infrared Imager (SEVIRI) are 4 km/15 min and 3 km/15 min, respectively. FengYun-4B (FY-4B) is particularly suitable for real-time precipitation monitoring over Asia due to its optimal geostationary orbit at 105°E. While other satellites provide high spatial and temporal resolution, FY-4B's longitudinal positioning offers unique observational advantages for the Asian region, which holds critical global importance across economic, geopolitical, cultural, and environmental dimensions.

\begin{table*}[!ht]
\centering
\caption{The main parameters of currently operational geostationary meteorological satellites from different regions as of January 1st, 2025.}
\label{tab:satellite} % 调整标签以区分原表格（可选）
\begin{tabular}{
p{1.8cm}<{\centering}  % 卫星型号列（原表头）
p{1.6cm}<{\centering}  % Locations列（原第一行参数）
p{1.6cm}<{\centering}  % Resolutions列（原第二行参数）
p{1.6cm}<{\centering}  % Payloads列（原第三行参数）
p{1.6cm}<{\centering}  % Bands列（原第四行参数）
p{1.6cm}<{\centering}  % Regions列（原第五行参数）
}
\toprule
% 新表头：原卫星型号转为第一列的表头，原参数转为列标题
\tabincell{c}{Satellites}
&\tabincell{c}{Locations}
&\tabincell{c}{Resolutions}
&\tabincell{c}{Payloads}
&\tabincell{c}{Bands}
&\tabincell{c}{Regions}
\\
\midrule[0.75pt]
% 每行对应原表格的一颗卫星，数据按“位置→分辨率→载荷→波段→所属区域”顺序排列
\tabincell{c}{FengYun-4B}
&\tabincell{c}{105°E}
&\tabincell{c}{4km/15min}
&\tabincell{c}{AGRI}
&\tabincell{c}{15}
&\tabincell{c}{China}
\\
\tabincell{c}{GOES-16}
&\tabincell{c}{75.2°W}
&\tabincell{c}{2km/10min}
&\tabincell{c}{ABI}
&\tabincell{c}{16}
&\tabincell{c}{USA}
\\
\tabincell{c}{GOES-18}
&\tabincell{c}{137°W}
&\tabincell{c}{2km/10min}
&\tabincell{c}{ABI}
&\tabincell{c}{16}
&\tabincell{c}{USA}
\\
\tabincell{c}{Himawari-9}
&\tabincell{c}{140.7°E}
&\tabincell{c}{2km/10min}
&\tabincell{c}{AHI}
&\tabincell{c}{16}
&\tabincell{c}{Japan}
\\
\tabincell{c}{Meteosat-11}
&\tabincell{c}{9.5°E}
&\tabincell{c}{3km/15min}
&\tabincell{c}{SEVIRI}
&\tabincell{c}{11}
&\tabincell{c}{Europe}
\\
\tabincell{c}{Meteosat-12}
&\tabincell{c}{0°}
&\tabincell{c}{2km/10min}
&\tabincell{c}{FCI}
&\tabincell{c}{16}
&\tabincell{c}{Europe}
\\
\bottomrule
\end{tabular}
\end{table*}
\subsection{IMERG}
The Integrated Multi-satellite Retrievals for GPM (IMERG) is a Level-3 product from the Global Precipitation Measurement (GPM) mission~\citep{huffman2015nasa,kidd2020global,tan2019imerg,huffman2023imerg}, a joint project of NASA and JAXA. It combines data from a constellation of satellites to provide the most advanced satellite-based precipitation estimates available. We use the IMERG V07B dataset, which offers three products with different latencies: Early Run (4-6 hours), Late Run (12 hours), and Final Run (3.5 months). The data has a spatial resolution of 0.1° and a temporal resolution of 30 minutes.

\begin{figure}[h!]
\centering
\centerline{\includegraphics[width=1\textwidth]{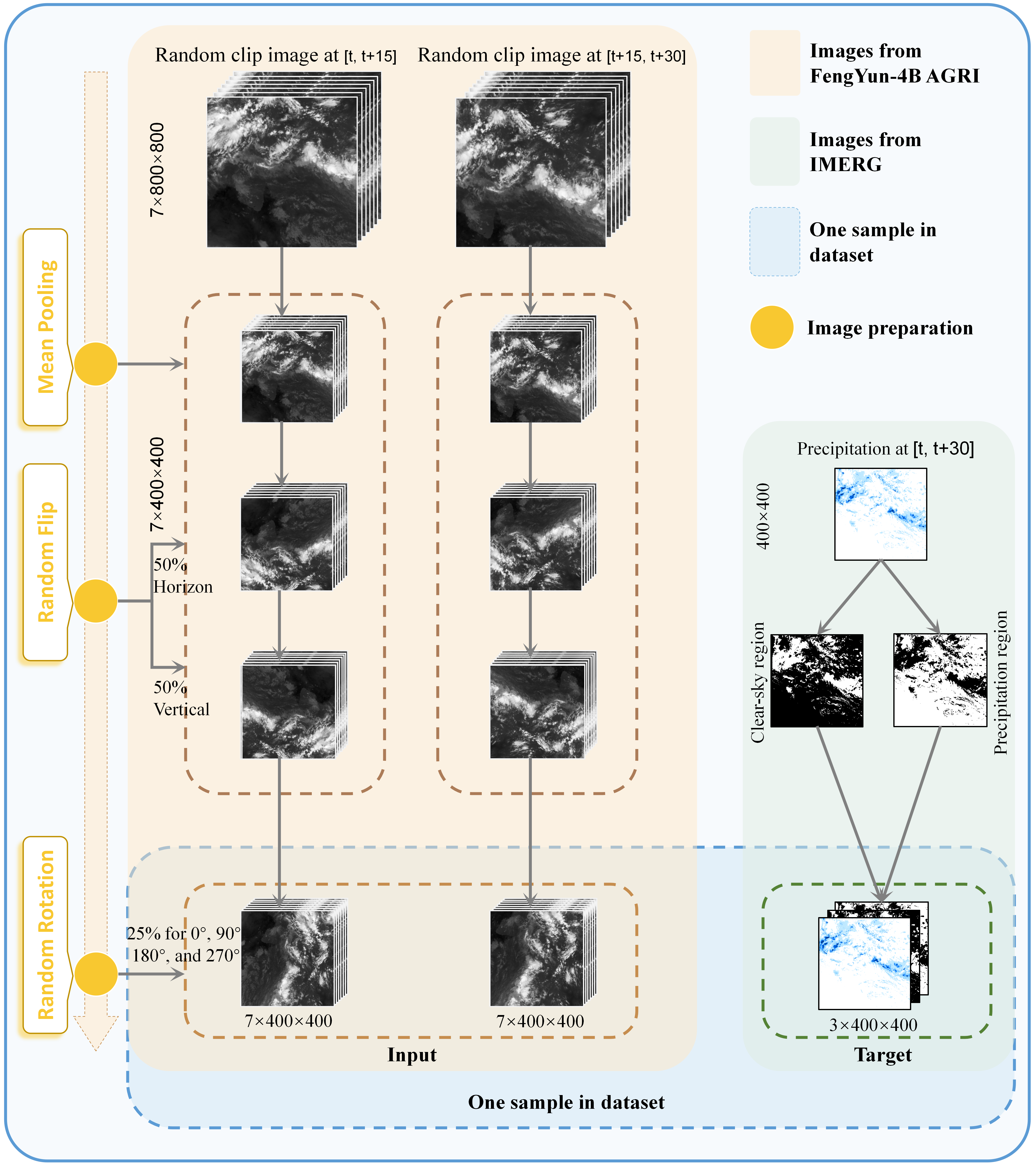}}
\caption{The imagery after random clipping, processing a 2×2 mean pooling to upscale before random flip and rotation. There is a 50\% probability that the subsequent image processing will occur in both the horizontal and vertical directions, respectively. The probability of every 90° rotation (0°, 90°, 180°, and 270°) is 25\%. The training and validation pair is a 17×400×400 feature map that contains two consecutive FY-4B imagery ((2×7)×400×400), precipitation, clean-sky region, and precipitation region (3×400×400).}
\label{fig:data}
\end{figure}

\begin{figure}[!ht]
\centering
\centerline{\includegraphics[width=1\textwidth]{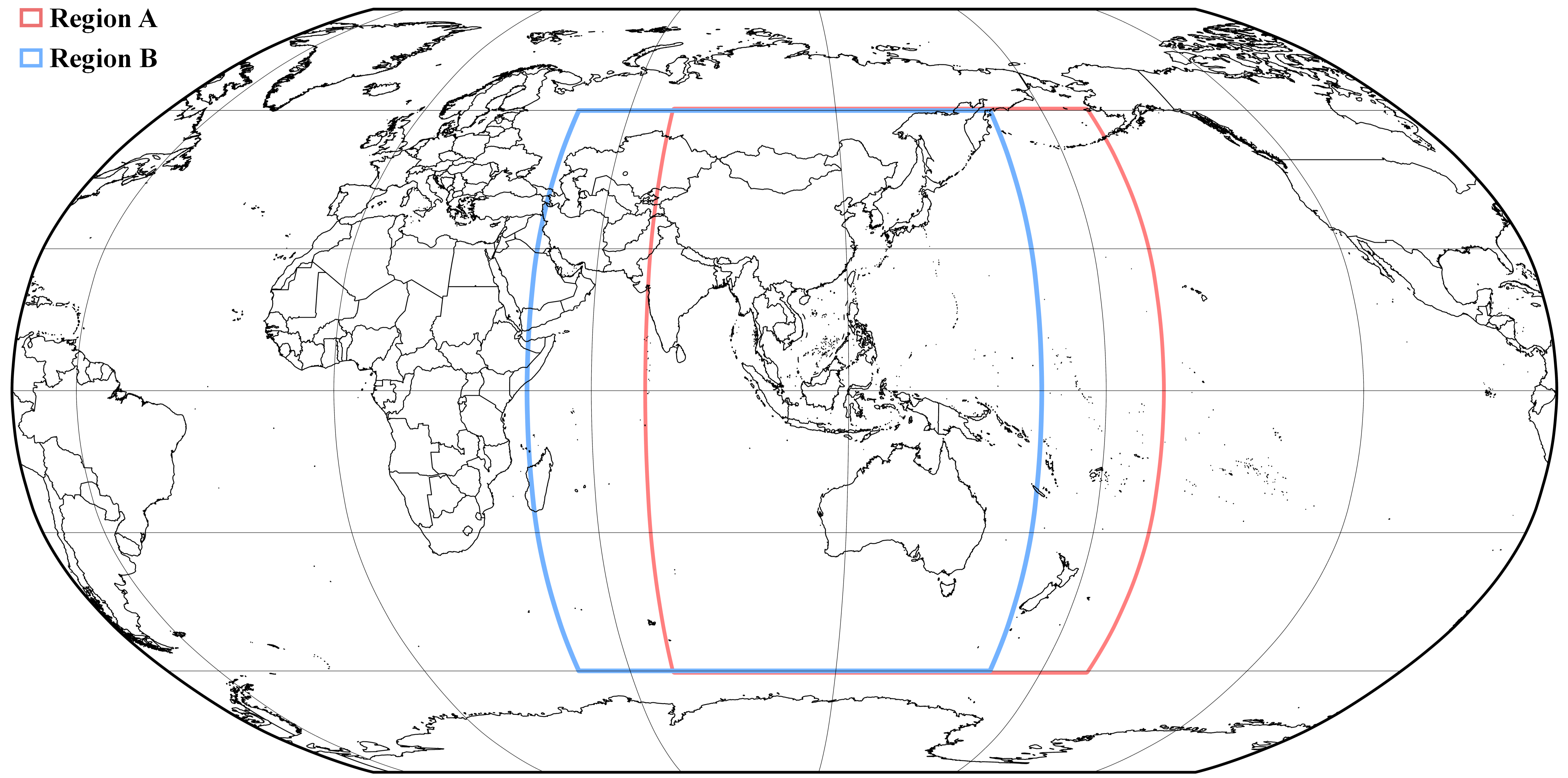}}
\caption{The region spans 120° for both longitude and latitude of FY-4B after recalibration~\citep{song5168620geoattx}. Before 31/1/2024, the middle longitude is 133°E (Region A) and after that interval, is 105°E (Region B).}
\label{fig:region}
\end{figure}

\subsection{FY-4B}
FengYun-4 (FY-4)~\citep{yang2017introducing} is the second generation of China's geostationary meteorological satellite series. FY-4B, the second of its series, was launched on June 3rd, 2021, and is an operational satellite. The main payloads onboard FY-4B are Advanced Geostationary Radiation Imager (AGRI), Geostationary Interferometric Infrared Sounder (GIIRS), Geostationary High-speed Imager (GHI), and Space Environment Package (SEP). This satellite has been extensively utilized across various research domains, including sea surface temperature quality assessment~\cite{he2024quality}, quantitative precipitation estimation~\cite{ma2022fy4qpe}, cloud image extrapolation~\cite{song5168620geoattx}, and atmospheric motion vector analysis~\cite{xia2024minute}. The AGRI instrument onboard FY-4B provides full-disk imagery at 15-minute intervals, which can be cropped to generate a 2,400×2,400 pixel image encompassing a 120° range in both longitude (45° E to 165° E) and latitude (60° S to 60° N) with a spatial resolution of 0.05°.

Huayu uses the seven water vapor channels (bands 9–15, 6.25–13.3 $\mu m$) from the FY-4B/AGRI instrument for real-time precipitation retrieval. The model was trained on IMERG data from 2023–2024 and validated on data from July–December 2022. The study region is shown in Fig.~\ref{fig:region}; note that the satellite's central longitude shifted from 133°E to 105°E after January 31, 2024. To prepare the dataTo prepare the data, original 2,746×2,746 full-disk images (4km resolution) were recalibrated into 2,746×2,746 (0.05° resolution) following~\cite{song5168620geoattx}, and randomly cropped into 800×800 patches. Each image was randomly cropped into nine 800×800 patches. And each patch went through random clipping, flipping, and rotation. To reconcile the different spatial resolutions of FY-4B (0.05°) and IMERG (0.1°), these patches were then mean-pooled to 400×400, as illustrated in Fig.~\ref{fig:data}. Training data were subsampled every 6 hours to ensure temporal diversity, while validation data were sampled every 49 hours. Samples with less than 2\% rainy area were filtered to accelerate convergence, resulting in a final dataset of 23,165 training and 728 validation samples (see the \href{https://github.com/songzijiang/GeoAttX}{source code repository} for details).

FY-4B QPE (FengYun-4B Quantitative Precipitation Estimate, 15-minute intermediate version) is a level 2 precipitation product from AGRI onboard the FY-4B. The original spatial-temporal resolution is 4km/15min at the subpoint, the same as full disk imagery.
\subsection{HadISD}
HadISD \citep{dunn2014pairwise,dunn2012hadisd,smith2011integrated,dunn2016expanding} is a global sub-daily dataset provided by the Met Office based on NOAA's ISD dataset after a suite of quality control tests on the major climatological variables. The version used in this work was v3.4.3.2025f, spans 1/1/1931 to 29/8/2025, and contains about 8,500 stations around the globe. In this work, we chose the 3-hour precipitation depth in the research area spanning 1/7/2022 to 31/12/2022 from 444 stations.

\subsection{PERSIANN and PERSIANN-CCS}
PERSIANN (Precipitation Estimation from Remotely Sensed Information using Artificial Neural Networks) is a current operational precipitation system with spatial resolution 0.25° based on geostationary satellite infrared imagery~\citep{nguyen2018persiann}. The Cloud Classification System of it (PERSIANN-CCS) is a real-time satellite precipitation high-resolution (0.04°) satellite precipitation product used widely, considered as the state-of-the-science precipitation data derived from IR observations from geostationary satellite~\citep{zhu2024pecam,hong2004precipitation,wang2020infrared,wang2021infrared}.
\subsection{CMORPH}
The Bias-Corrected Climate Prediction Center Morphing technique (CMORPH) is a NOAA (National Oceanic and Atmospheric Administration) Climate Data Record (CDR) using passive microwave and infrared sensors aboard multiple satellites~\citep{joyce2004cmorph}. The resolution of it is 8km/30min with a latency of 3-4 months.
\subsection{GSMaP\_NOW}
Global Rainfall Map Realtime version (GSMaP\_NOW) is produced by JAXA~\citep{kubota2020global}. GSMaP\_NOW contains two versions, depending on whether calibrated by gauges. The gauge-calibrated rainfall product was used in this work.

%% file: content/table/shanghaigauge.tex
\begin{table*}[!t]
\setlength{\extrarowheight}{2pt}
  % \tiny
  \scriptsize
  \centering
  \caption{The comparison POD, FAR, ACC, and CSI between IMERG Early Run (ER) and Huayu (HY) when evaluated on 28 gauge stations in Shanghai using precipitation rate (mm/hr). The better results are highlighted in bold.}
  \label{tab:comparison}
  % \resizebox{\textwidth}{!}{
  \begin{tabular}{
  p{2.1cm}<{\centering}
  p{2.1cm}<{\centering}
  p{1.4cm}<{\centering}
  p{1.4cm}<{\centering}
  p{1.4cm}<{\centering}
  p{1.4cm}<{\centering}
  } \toprule
 \tabincell{c}{\textbf{Station ID}}
&\tabincell{c}{\textbf{Lon./Lat.}}
  % &\tabincell{c}{$\mathbf{RMSE}\uparrow$}
&\tabincell{c}{{$\mathbf{POD}\uparrow$}\\\hline{ER/\textbf{HY}}}
&\tabincell{c}{{$\mathbf{FAR}\downarrow$}\\\hline{ER/\textbf{HY}}}
&\tabincell{c}{{$\mathbf{ACC}\uparrow$}\\\hline{ER/\textbf{HY}}}
&\tabincell{c}{{$\mathbf{CSI}\uparrow$}\\\hline{ER/\textbf{HY}}}
\\
\midrule
% \midrule[0.75pt0.75pt]

\tabincell{c}{\textbf{SW62701402}} &\tabincell{c}{121.189/31.761} &\tabincell{c}{$ 0.88$/$\mathbf{0.98 }$}&\tabincell{c}{$ 0.42$/$\mathbf{0.44 }$}&\tabincell{c}{$\mathbf{ 0.67}$/$0.66 $}&\tabincell{c}{$ 0.54$/$\mathbf{0.56 }$}\\
\tabincell{c}{\textbf{SW62701500}} &\tabincell{c}{121.606/31.522} &\tabincell{c}{$ 0.90$/$\mathbf{0.99 }$}&\tabincell{c}{$\mathbf{ 0.39}$/$0.37 $}&\tabincell{c}{$ 0.66$/$\mathbf{0.70 }$}&\tabincell{c}{$ 0.57$/$\mathbf{0.63 }$}\\
\tabincell{c}{\textbf{SW62724635}} &\tabincell{c}{121.407/31.613} &\tabincell{c}{$ 0.88$/$\mathbf{0.98 }$}&\tabincell{c}{$ 0.43$/$\mathbf{0.44 }$}&\tabincell{c}{$ 0.64$/$0.64 $}&\tabincell{c}{$ 0.53$/$\mathbf{0.55 }$}\\
\tabincell{c}{\textbf{SW63401100}} &\tabincell{c}{121.237/30.964} &\tabincell{c}{$ 0.86$/$\mathbf{0.94 }$}&\tabincell{c}{$ 0.42$/$\mathbf{0.44 }$}&\tabincell{c}{$\mathbf{ 0.61}$/$0.59 $}&\tabincell{c}{$ 0.53$/$\mathbf{0.54 }$}\\
\tabincell{c}{\textbf{SW63401500}} &\tabincell{c}{121.488/31.244} &\tabincell{c}{$ 0.91$/$\mathbf{0.97 }$}&\tabincell{c}{$ 0.31$/$\mathbf{0.34 }$}&\tabincell{c}{$\mathbf{ 0.71}$/$0.70 $}&\tabincell{c}{$ 0.64$/$\mathbf{0.65 }$}\\
\tabincell{c}{\textbf{SW63401750}} &\tabincell{c}{121.499/31.382} &\tabincell{c}{$ 0.87$/$\mathbf{0.97 }$}&\tabincell{c}{$\mathbf{ 0.44}$/$0.42 $}&\tabincell{c}{$ 0.60$/$\mathbf{0.64 }$}&\tabincell{c}{$ 0.52$/$\mathbf{0.57 }$}\\
\tabincell{c}{\textbf{SW63402200}} &\tabincell{c}{121.212/30.929} &\tabincell{c}{$ 0.84$/$\mathbf{0.93 }$}&\tabincell{c}{$ 0.37$/$\mathbf{0.39 }$}&\tabincell{c}{$ 0.64$/$0.64 $}&\tabincell{c}{$ 0.56$/$\mathbf{0.59 }$}\\
\tabincell{c}{\textbf{SW63402500}} &\tabincell{c}{121.003/30.903} &\tabincell{c}{$ 0.82$/$\mathbf{0.93 }$}&\tabincell{c}{$ 0.38$/$0.38 $}&\tabincell{c}{$ 0.61$/$\mathbf{0.64 }$}&\tabincell{c}{$ 0.55$/$\mathbf{0.60 }$}\\
\tabincell{c}{\textbf{SW63402910}} &\tabincell{c}{121.035/30.961} &\tabincell{c}{$ 0.86$/$\mathbf{0.93 }$}&\tabincell{c}{$ 0.36$/$\mathbf{0.39 }$}&\tabincell{c}{$\mathbf{ 0.65}$/$0.63 $}&\tabincell{c}{$ 0.58$/$0.58 $}\\
\tabincell{c}{\textbf{SW63403100}} &\tabincell{c}{121.047/31.052} &\tabincell{c}{$ 0.88$/$\mathbf{0.94 }$}&\tabincell{c}{$ 0.33$/$\mathbf{0.35 }$}&\tabincell{c}{$\mathbf{ 0.67}$/$0.66 $}&\tabincell{c}{$ 0.61$/$\mathbf{0.62 }$}\\
\tabincell{c}{\textbf{SW63403190}} &\tabincell{c}{121.034/31.024} &\tabincell{c}{$ 0.85$/$\mathbf{0.93 }$}&\tabincell{c}{$ 0.35$/$\mathbf{0.36 }$}&\tabincell{c}{$ 0.64$/$0.64 $}&\tabincell{c}{$ 0.59$/$\mathbf{0.61 }$}\\
\tabincell{c}{\textbf{SW63403200}} &\tabincell{c}{120.897/31.020} &\tabincell{c}{$ 0.84$/$\mathbf{0.93 }$}&\tabincell{c}{$ 0.32$/$\mathbf{0.33 }$}&\tabincell{c}{$ 0.66$/$\mathbf{0.67 }$}&\tabincell{c}{$ 0.60$/$\mathbf{0.63 }$}\\
\tabincell{c}{\textbf{SW63403800}} &\tabincell{c}{120.917/31.116} &\tabincell{c}{$ 0.86$/$\mathbf{0.95 }$}&\tabincell{c}{$ 0.33$/$\mathbf{0.35 }$}&\tabincell{c}{$ 0.67$/$0.67 $}&\tabincell{c}{$ 0.60$/$\mathbf{0.63 }$}\\
\tabincell{c}{\textbf{SW63404000}} &\tabincell{c}{121.118/31.117} &\tabincell{c}{$ 0.84$/$\mathbf{0.93 }$}&\tabincell{c}{$\mathbf{ 0.33}$/$0.32 $}&\tabincell{c}{$ 0.64$/$\mathbf{0.68 }$}&\tabincell{c}{$ 0.59$/$\mathbf{0.64 }$}\\
\tabincell{c}{\textbf{SW63404100}} &\tabincell{c}{121.179/31.094} &\tabincell{c}{$ 0.84$/$\mathbf{0.92 }$}&\tabincell{c}{$ 0.32$/$0.32 $}&\tabincell{c}{$ 0.65$/$\mathbf{0.68 }$}&\tabincell{c}{$ 0.60$/$\mathbf{0.64 }$}\\
\tabincell{c}{\textbf{SW63404540}} &\tabincell{c}{121.883/31.007} &\tabincell{c}{$ 0.93$/$\mathbf{0.98 }$}&\tabincell{c}{$ 0.45$/$\mathbf{0.46 }$}&\tabincell{c}{$\mathbf{ 0.61}$/$0.60 $}&\tabincell{c}{$ 0.53$/$0.53 $}\\
\tabincell{c}{\textbf{SW63404554}} &\tabincell{c}{121.641/31.016} &\tabincell{c}{$ 0.88$/$\mathbf{0.94 }$}&\tabincell{c}{$\mathbf{ 0.39}$/$0.38 $}&\tabincell{c}{$ 0.62$/$\mathbf{0.64 }$}&\tabincell{c}{$ 0.56$/$\mathbf{0.60 }$}\\
\tabincell{c}{\textbf{SW63404590}} &\tabincell{c}{121.679/31.317} &\tabincell{c}{$ 0.90$/$\mathbf{0.98 }$}&\tabincell{c}{$ 0.38$/$0.38 $}&\tabincell{c}{$ 0.66$/$\mathbf{0.67 }$}&\tabincell{c}{$ 0.58$/$\mathbf{0.61 }$}\\
\tabincell{c}{\textbf{SW63404595}} &\tabincell{c}{121.570/31.114} &\tabincell{c}{$ 0.90$/$\mathbf{0.96 }$}&\tabincell{c}{$ 0.34$/$0.34 $}&\tabincell{c}{$ 0.67$/$\mathbf{0.68 }$}&\tabincell{c}{$ 0.62$/$\mathbf{0.64 }$}\\
\tabincell{c}{\textbf{SW63404855}} &\tabincell{c}{121.774/31.063} &\tabincell{c}{$ 0.91$/$\mathbf{0.96 }$}&\tabincell{c}{$ 0.42$/$\mathbf{0.43 }$}&\tabincell{c}{$\mathbf{ 0.62}$/$0.61 $}&\tabincell{c}{$ 0.55$/$0.55 $}\\
\tabincell{c}{\textbf{SW63404905}} &\tabincell{c}{121.702/31.198} &\tabincell{c}{$ 0.91$/$\mathbf{0.96 }$}&\tabincell{c}{$ 0.35$/$0.35 $}&\tabincell{c}{$ 0.67$/$\mathbf{0.69 }$}&\tabincell{c}{$ 0.61$/$\mathbf{0.64 }$}\\
\tabincell{c}{\textbf{SW63405000}} &\tabincell{c}{121.063/31.269} &\tabincell{c}{$ 0.83$/$\mathbf{0.94 }$}&\tabincell{c}{$ 0.39$/$\mathbf{0.42 }$}&\tabincell{c}{$\mathbf{ 0.64}$/$0.62 $}&\tabincell{c}{$ 0.54$/$\mathbf{0.56 }$}\\
\tabincell{c}{\textbf{SW63405460}} &\tabincell{c}{121.334/31.417} &\tabincell{c}{$ 0.86$/$\mathbf{0.95 }$}&\tabincell{c}{$\mathbf{ 0.39}$/$0.38 $}&\tabincell{c}{$ 0.64$/$\mathbf{0.67 }$}&\tabincell{c}{$ 0.55$/$\mathbf{0.60 }$}\\
\tabincell{c}{\textbf{SW63405480}} &\tabincell{c}{121.246/31.376} &\tabincell{c}{$ 0.84$/$\mathbf{0.92 }$}&\tabincell{c}{$\mathbf{ 0.38}$/$0.37 $}&\tabincell{c}{$ 0.63$/$\mathbf{0.66 }$}&\tabincell{c}{$ 0.56$/$\mathbf{0.60 }$}\\
\tabincell{c}{\textbf{SW63405800}} &\tabincell{c}{121.845/30.840} &\tabincell{c}{$ 0.90$/$\mathbf{0.95 }$}&\tabincell{c}{$ 0.45$/$\mathbf{0.49 }$}&\tabincell{c}{$\mathbf{ 0.63}$/$0.57 $}&\tabincell{c}{$\mathbf{ 0.52}$/$0.49 $}\\
\tabincell{c}{\textbf{SW63405900}} &\tabincell{c}{121.367/30.730} &\tabincell{c}{$ 0.89$/$\mathbf{0.97 }$}&\tabincell{c}{$ 0.40$/$\mathbf{0.43 }$}&\tabincell{c}{$\mathbf{ 0.66}$/$0.63 $}&\tabincell{c}{$ 0.56$/$0.56 $}\\
\tabincell{c}{\textbf{SW63425700}} &\tabincell{c}{121.748/31.120} &\tabincell{c}{$ 0.93$/$\mathbf{0.96 }$}&\tabincell{c}{$ 0.35$/$\mathbf{0.37 }$}&\tabincell{c}{$\mathbf{ 0.69}$/$0.66 $}&\tabincell{c}{$\mathbf{ 0.62}$/$0.61 $}\\
\tabincell{c}{\textbf{SW63439055}} &\tabincell{c}{121.673/31.156} &\tabincell{c}{$ 0.90$/$\mathbf{0.96 }$}&\tabincell{c}{$\mathbf{ 0.35}$/$0.34 $}&\tabincell{c}{$ 0.66$/$\mathbf{0.69 }$}&\tabincell{c}{$ 0.61$/$\mathbf{0.64 }$}\\
\midrule
\tabincell{c}{\textbf{\scriptsize{Overall}}} &\tabincell{c}{\textbf{--/--}} &\tabincell{c}{$ 0.87$/$\mathbf{0.95 }$}&\tabincell{c}{$ 0.38$/$\mathbf{0.39 }$}&\tabincell{c}{$ 0.65$/$0.65 $}&\tabincell{c}{$ 0.57$/$\mathbf{0.60 }$}\\

 \bottomrule
 \end{tabular}
 % }
 \label{tab:stations}
\end{table*}